\documentclass[acmsmall,screen,nonacm]{acmart}

\theoremstyle{definition}

\usepackage{mathtools}
\usepackage{amsmath}
\usepackage{amsthm}
\usepackage{mathpartir}
\usepackage{proof}

\usepackage{booktabs}
\usepackage{multirow}
\usepackage{siunitx}
\usepackage{tabularx}

\usepackage{multicol}
\usepackage{paralist}
\usepackage{enumitem}

\usepackage{xcolor}
\usepackage{tikz}
\usetikzlibrary{fit}
\usetikzlibrary{arrows,shapes,positioning,shadows,trees}

\usepackage[linguistics]{forest}

\usepackage{caption}
\usepackage{subcaption}
\usepackage{wrapfig}

\usepackage{comment}
\usepackage{mdframed}
\usepackage{lineno}
\usepackage{xspace}

\usepackage{listings}

\definecolor{keywordblue}{RGB}{0,76,153}
\definecolor{typepurple}{RGB}{120,60,140}
\definecolor{commentgray}{RGB}{90,90,90}

\lstdefinestyle{code-legacy}{
  language=C,
  basicstyle=\ttfamily\small,
  columns=fullflexible,
  framesep=6pt,
  frame=single,
  numbers=left,
  numberstyle=\scriptsize\color{commentgray},
  numbersep=6pt,
  breaklines=true,
  xleftmargin=12pt,
  framexleftmargin=8pt,
  keywordstyle=\color{keywordblue}\bfseries,
  commentstyle=\color{commentgray}\itshape,
  stringstyle=\color{typepurple},
  morekeywords={
    Lgetstack,Lsetstack,Lop,Lload,Lstore,
    Lcall,Ltailcall,Llabel,Lgoto,Lcond,
    Ljumptable,Lbuiltin,Lreturn,
    Mgetstack,Msetstack,Mgetparam,Mop,
    Mload,Mstore,Mcall,Mtailcall,
    Mlabel,Mgoto,Mcond,Mjumptable,
    Mbuiltin,Mreturn
  },
  morekeywords=[2]{
    Local,Incoming,Outgoing,
    Tint,Tlong,Tfloat,Tsingle
  },
  keywordstyle=[2]\color{typepurple}
}

\definecolor{codeink}{RGB}{35,52,65}
\definecolor{coderule}{gray}{0.60}

\lstdefinestyle{code}{
  style=code-legacy,
  basicstyle=\ttfamily\footnotesize,
  columns=fixed,
  basewidth=0.5em,
  keepspaces=true,
  tabsize=2,
  showstringspaces=false,
  frame=tb,
  framerule=0.3pt,
  rulecolor=\color{coderule},
  framesep=3pt,
  framexleftmargin=0pt,
  framexrightmargin=0pt,
  xleftmargin=0pt,
  xrightmargin=0pt,
  numbers=none,
  aboveskip=4pt,
  belowskip=4pt,
  lineskip=-1pt,
  breakatwhitespace=true,
  breakindent=1em,
  keywordstyle=\color{codeink}\bfseries,
  keywordstyle=[2]\color{codeink},
  commentstyle=\color{commentgray}\itshape,
  stringstyle=\color{black},
  morekeywords={call,tailcall,skip,locals,stack,var,parameters}
}
\lstdefinestyle{code-asm}{
  style=code,
  language={},
  morekeywords={mov,bl,cbnz,ldr,b,add,ret,stp},
  keywordstyle=\color{codeink}\bfseries
}
\lstdefinestyle{code-facts}{
  style=code,
  language={},
  morekeywords={FUNCTION,TYPE},
  keywordstyle=\color{codeink}\bfseries
}
\newcommand{\codeheading}[1]{%
  \par\raggedright{\small\bfseries #1\par}\nobreak
}
\newcommand{\codegap}{\hspace{1.2em}}

\usepackage{coqdoc}

\usepackage{hyperref}

\hypersetup{
  colorlinks=true,
  linkcolor={red!70!black},
  citecolor={red!70!black},
  urlcolor={blue!70!black}
}

\definecolor{ao(english)}{rgb}{0.0,0.5,0.0}

\setlist[description]{leftmargin=\parindent}
\setlist[itemize]{left=.618\parindent .. 1.618\parindent}

\newif\ifdiff
\difffalse

\newcommand{\tool}{\textsc{DeCompCert}\xspace}

\DeclareRobustCommand{\ccname}[1]{%
  \texorpdfstring{\ifmmode\mathsf{#1}\else\textsf{#1}\fi}{#1}%
}

\DeclareRobustCommand{\rq}[1]{%
  \texorpdfstring{\textsf{\textbf{RQ#1}}}{RQ#1}%
}
\newenvironment{researchquestions}{%
  \begin{description}[font=\normalfont,align=left,leftmargin=3.6em,
    labelwidth=2.8em,labelsep=0.8em,topsep=4pt,itemsep=3pt,parsep=0pt]
}{%
  \end{description}
}
\newcommand{\rqquestion}[2]{\item[\rq{#1}]\textbf{#2}}

\DeclareUnicodeCharacter{00B7}{\cdot}
\DeclareUnicodeCharacter{00D7}{\ifmmode\times\else\texttimes\fi}
\DeclareUnicodeCharacter{03B3}{\ensuremath{\gamma}}
\DeclareUnicodeCharacter{03BA}{\ensuremath{\kappa}}
\DeclareUnicodeCharacter{03C4}{\ensuremath{\tau}}
\DeclareUnicodeCharacter{2081}{\ensuremath{{}_1}}
\DeclareUnicodeCharacter{2082}{\ensuremath{{}_2}}
\DeclareUnicodeCharacter{2113}{\ensuremath{\ell}}
\DeclareUnicodeCharacter{21A6}{\ensuremath{\mapsto}}
\DeclareUnicodeCharacter{21D2}{\implies}
\DeclareUnicodeCharacter{21D3}{\Downarrow}
\DeclareUnicodeCharacter{2200}{\forall}
\DeclareUnicodeCharacter{2203}{\exists}
\DeclareUnicodeCharacter{2208}{\in}
\DeclareUnicodeCharacter{2218}{\ensuremath{\circ}}
\DeclareUnicodeCharacter{2227}{\land}
\DeclareUnicodeCharacter{222A}{\cup}
\DeclareUnicodeCharacter{2264}{\leqslant}
\DeclareUnicodeCharacter{2291}{\sqsubseteq}

\newcommand{\rname}[1]{\ensuremath{[\textsc{#1}]}}
\let\xinferrule\inferrule
\renewcommand{\inferrule}[3][]{%
	\ifx&#1&
	\xinferrule*{#2}{#3}
	\else
	\xinferrule*[right=\rname{#1}]{#2}{#3}
	\fi}

\def\substitute#1by#2{[#2/#1]}

\newcommand*\vtrue{\ensuremath{{t\kern-.4ex t}}}
\newcommand*\vfalse{\ensuremath{{f\kern-.7ex f}}}

\newcommand{\tse}{\tau_{s}}

\newcommand{\ltconst}[1]{C(\tse)} 

\begin{document}

\title{DeCompCert: A Verified Decompiler for CompCert-Generated AArch64}

\author{Swarn Priya}
\affiliation{%
  \institution{Washington State University}
  \country{USA}}

\author{Basavesh Ammanaghatta Shivakumar}
\affiliation{%
  \institution{Washington State University}
  \country{USA}}

\author{Thomas Gilray}
\affiliation{%
  \institution{Washington State University}
  \country{USA}}

\author{Kristopher Micinski}
\affiliation{%
  \institution{Syracuse University}
  \country{USA}}

\begin{abstract}
Decompilation recovers high-level code from object- or assembly-level code and
is foundational to reverse engineering, binary hardening, and vulnerability
discovery. In practice, decompilation is subject to counterveiling tensions of soundness and transparency: techniques that capture semantics exactly will maintain some source abstractions, which represent obfuscations in synthesized high-level code. To minimize this, production tools such as IDA and Ghidra rely on sophisticated analyses and heuristics, but that come without
a formal end-to-end soundness guarantee.
We present \tool{}, a Rocq-verified decompiler for the AArch64 assembly generated by CompCert, and focus on recovery and analysis of CompCert-built artifacts whose source is unavailable, rather than decompilation of arbitrary assembly. 
\tool{} reconstructs CompCert \ccname{Csyntax}
through nine stages using both directly verified reconstruction and translation validation. Untrusted OCaml and Datalog-based analyses propose candidates, but provably-correct validators accept only candidates that are provably equivalent to a lower-level input.
For every reconstruction accepted from a typed CompCert \ccname{Asm} program,
the composed theorem guarantees that every (well defined, safe) observable
behavior of the reconstructed program under CompCert's \ccname{Cstrategy}
semantics is matched by the \ccname{Asm} program.

We evaluate \tool{} on CompCert-generated assembly from approximately
2.16 million lines of C and demonstrate full reverse-pipeline completeness
when exact function and call signatures are provided. Under this configuration,
\tool{} reconstructs and recompiles all 3,173 eligible contexts under both
CompCert's unoptimized and optimized pipelines. Thus, when exact function and
call signatures are available, \tool{} achieves 100\% connected end-to-end
decompilation coverage over the 3,173 eligible contexts. Without retained
signatures, a
cumulative portfolio combining checked hint consensus, reconstruction
improvements, and longer bounded searches finds complete validator-checked
chains for 2,606 of 3,164 contexts (82.4\%). The primary platform-plus-IDA
configuration accepts 2,407 contexts (76.1\%). Failed searches produce no
certified C result. Native-execution and output-fidelity studies further
exercise the emitted C and characterize the remaining limitations in signature
recovery, source resemblance, and optimized signature-free reconstruction.
Across 45,660 matched function instances, we also measure source quality and
show that Datalog-guided sharing of compatible memory layouts reduces casts and
improves recovery of physical field layouts without weakening validation.
\end{abstract}

\maketitle

\section{Introduction}

Decompilation is the process of inferring source code from low-level (typically
object- or assembly-level) code. It plays a key role in reverse engineering,
binary analysis, and instrumentation~\cite{cifuentes1994reverse,verbeek2020foxdec}. In this work,
we study a precisely scoped, useful
instance of this problem: recovering C from AArch64 artifacts produced by the verified CompCert compiler. Such recovery supports auditing deployed
CompCert-built software, inspecting certified artifacts whose source is no
longer available, and reusing source-level analyses or instrumentation.
Reasoning directly about assembly is challenging because compilation removes
much of the structure and semantic information originally present in the source program.
For example, function prototypes are stripped; the calling convention identifies possible argument and result registers,
but does not tell us whether their values are integers, pointers, or
parts of higher-level aggregates. Some constructs, such as variable names, simply cannot be recovered; thus, \emph{any} decompilation is essentially a representative of some equivalence class of source-level terms. For the CompCert-generated inputs considered here, decompilation is \emph{sound} when every defined observable behavior of the synthesized source term is exhibited by the supplied typed CompCert \ccname{Asm} program.
Production decompilers such as IDA Pro~\cite{idapro} and
Ghidra~\cite{ghidra} do not provide end-to-end soundness guarantees, and
empirical studies have identified semantically incorrect
outputs~\cite{liu2020decompiler,298204}. The synthesized high-level code often requires cleanup via an iterative process of manual fixups (to types, variable names, etc.). While this may suffice for reading the code, it precludes scenarios where the decompiled program is used as the basis for subsequent analysis, instrumentation, or hardening.

In this paper, we present, to our knowledge, the first end-to-end sound
decompilation of CompCert-generated AArch64 to C via translation
validation~\cite{pnueli1998translation,necula2000translation}. The essence of
our approach (\S~\ref{sec:decompilation}) is to separate generation and
checking. \tool{} uses CompCert's intermediate representations
~\cite{10.1145/1538788.1538814} to give semantics to progressively higher-level
reconstructions. Starting with \ccname{Asm}, we reconstruct the inputs of
CompCert's passes using untrusted lifters with certified validators or directly
verified Rocq reconstruction~\cite{rocqmanual}. Other transformations require
search, for which unsound Datalog-based analyses~\cite{ceri1989datalog} propose candidates
that validators proved sound in Rocq subsequently accept.
\tool{} then works by composing each of the certified passes together: 
as we lift to progressively higher-level IRs, each validated higher-level artifact becomes the input of the
next pass, ultimately yielding a \ccname{Csyntax} program which has been proven to simulate its assembly-level equivalent. 
\tool{} requires a typed, unexpanded CompCert \ccname{Asm} program; its
correctness theorem does not require source-derived function prototypes. The theorem
unconditionally provides a forward simulation from reconstructed
\ccname{Cstrategy} semantics to the input \ccname{Asm} artifact; when the program is safe, we also show exact
observational equivalence. 


Our evaluation deliberately separates two questions. Given correct function
and call signatures, \tool{} reconstructs, certifies, and recompiles all 3,173
eligible contexts across sixteen application groups and both CompCert
pipelines. This 100\% result measures reverse-pipeline completeness. Starting
instead from CompCert-emitted assembly without retained signatures, \tool{}
inference alone accepts 1,874 of 3,164 contexts (59.2\%). This lower rate
measures the additional challenge of recovering interfaces absent from
assembly text, not a failure of the verified reverse passes. Checked platform
proposals raise coverage to 2,218 (70.1\%); configurations adding checked
Ghidra or IDA proposals reach 2,325 (73.5\%) or 2,407 (76.1\%), respectively.
The cumulative portfolio reaches 2,606 (82.4\%) with additional reconstruction
improvements and longer runs. All configurations use the same validator
boundary. On the shared fourteen-application scope, \tool{}'s first inferred
signature matches 40.9\% of held-out CompCert signatures, compared with 28.6\%
for Ghidra. These measurements distinguish first-choice signature agreement
from complete chains accepted by the proved validators. A failed or exhausted
search produces no unchecked result.

\paragraph{Contributions.}
This paper makes the following contributions:

\begin{itemize}
    \item \textbf{Verified and compositional decompilation.}
    \tool{} reverses nine CompCert transformations using directly verified
    reconstruction and translation validation. We prove the correctness of
    both the individual passes and their complete composition. Whenever the
    whole decompiler accepts, a machine-checked theorem guarantees that every
    defined observable behavior of the reconstructed C program is matched by the
    supplied \ccname{Asm} program.

    \item \textbf{Inference-guided reconstruction.}
    Application-wide Datalog (Ascent~\cite{sahebolamri2022ascent,10.1145/3622840}) analyses recover stack and register relationships,
    types, layouts, and function-signature candidates. \tool{} can begin from
    assembly text without retained source signatures: complete typed-\ccname{Asm}
    assignments are proposed from register and stack accesses interpreted under the ABI,
    checked target-platform declarations, and independently bound Ghidra or IDA signatures,
    then passed through the unchanged proved validators. The same whole-decompiler theorem applies
    to every accepted reconstruction relative to its inferred typed \ccname{Asm} input.

    \item \textbf{Implementation and evaluation.}
    From CompCert-emitted assembly without retained signatures, \tool{} alone
    accepts 59.2\% of configured contexts; platform proposals raise coverage
    to 70.1\%, configurations adding Ghidra or IDA reach 73.5\% or 76.1\%, and
    the cumulative portfolio reaches 82.4\%. A separate exact-signature oracle
    baseline accepts all 3,173 eligible contexts at \texttt{-O0},
    \texttt{-O1}, and \texttt{-O2} (the latter two use the same CompCert
    optimized pipeline); this isolates reconstruction coverage and is not the
    operational input model.

    \item \textbf{Evaluation of reconstructed-code quality.}
    We measure source similarity across 45,660 matched function instances, compare a
    fixed subset with Ghidra, and evaluate how Ascent-guided type and layout
    recovery changes the reconstructed code.
\end{itemize}

\section{Formal Guarantees for Sound Decompilation}

\begin{figure}[tbp]
\centering
\captionsetup{skip=5pt}
\begin{minipage}[t]{0.36\linewidth}
\codeheading{C source}
\begin{lstlisting}[style=code,basicstyle=\ttfamily\footnotesize]
extern int compute(int);
extern int finish(int);
int process(int x) {
  int y = compute(x);
  if (x == 0) {
    return finish(y);
  } else {
    return x + y;
  }
}
\end{lstlisting}
\end{minipage}%
\codegap%
\begin{minipage}[t]{0.28\linewidth}
\codeheading{Assembly}
\begin{lstlisting}[style=code-asm,basicstyle=\ttfamily\footnotesize]
process:
    mov   x19, x0
    bl    _compute
    cbnz  w19, L1
    ldr   x19, [sp, #16]
    b     _finish
L1:
    add   w0, w19, w0
    ldr   x19, [sp, #16]
    ret   x30
\end{lstlisting}
\end{minipage}

\caption{Abridged AArch64 compilation of \textsf{process}. Stack-pointer
adjustments, the initial save of X19, return-address preservation, and CFI
directives are omitted to emphasize the data and control flow.}
\Description{Side-by-side C and abridged AArch64 for a function with a call,
a conditional, a tail call, and an addition.}
\label{fig:process-example}
\end{figure}
We explain our approach by way of example. 
Figure~\ref{fig:process-example} shows a small C function containing a call, a conditional, and two return paths.  Although the source is simple, compilation removes most of its explicit structure. The argument \textsf{x}, initially passed in register \textsf{x0}, is moved to the callee-saved register \textsf{x19} so that it remains available after the call to \textsf{compute}; the returned value \textsf{y} then occupies \textsf{x0}. These registers no longer carry the names or roles of the corresponding source variables. The source condition \textsf{x == 0} is represented by the inverted branch \textsf{cbnz w19, L1}; furthermore, \textsf{return finish(y)} is translated as \textsf{b \_finish}, which must be recognized as a tail call. The expression \(x+y\) survives only as an addition over physical registers. The complete assembly also contains stack-frame allocation and callee-save operations with no direct source-level counterparts~\cite{arm2024a64,arm2026aapcs64}.

Traditional decompilation is analysis-directed: heuristics reconstruct higher-level code informed by the results of a static analysis. As an example, consider the recovery of function prototypes from low-level code. In the assembly, 
function prototypes are absent, but propagating known information can help infer them: the use of \textsf{w19} in \textsf{cbnz} suggests a 32-bit argument to \texttt{process}; the subsequent addition uses the value returned in \texttt{w0} by \texttt{compute}, suggesting an integer result; and the tail branch to \texttt{finish} connects
the outgoing register state with another function interface.  Based on the result of these analyses, \tool{} infers a candidate decompilation consistent with the observed analysis facts. 
But crucially, \tool{} does \emph{not} require analyses or transformations to be sound: instead, \tool{} uses certified validators to establish a simulation between the low- and high-level IRs~\cite{compcertsmallstep}. For each pass, \tool{}'s validator either accepts or rejects the lifting from lower- to higher-level code. Using CompCert's semantics~\cite{10.1145/1538788.1538814}, we prove that, if the validator accepts, then the high-level fragment simulates its low-level equivalent.

\subsection{No Search Required: One-to-One Liftings}

Even when reversing, some compiler passes are functional in flavor; when low-level constructs correspond one-to-one with higher-level constructs, we proceed precisely as CompCert does: by writing functions in Rocq~\cite{rocqmanual}, with correctness established by proving a simulation between the source and reconstructed programs. Let $\mathcal{D}_{L\rightarrow H}$ be a reconstruction pass
that transforms a lower-level program $P_L$ into a
higher-level program $P_H$. We prove that 
reconstruction establishes a simulation:
\[
  \mathcal{D}_{L\rightarrow H}(P_L)=\mathsf{OK}(P_H)
  \Longrightarrow
  \mathsf{forward\_simulation}
    \bigl(\mathsf{Sem}_H(P_H),\mathsf{Sem}_L(P_L)\bigr).
\]

The proof defines a relation $\mathcal{R}$ between execution
states of the reconstructed and lower-level programs.
We establish three main properties:

\begin{enumerate}
\item \textbf{Initial-state correspondence.}
Every initial state of $P_H$ is related to an initial
state of $P_L$.

\item \textbf{Step simulation.}
For every pair of related states, an execution step of
$P_H$ can be matched by an execution sequence of $P_L$
with the same observable trace, preserving $\mathcal{R}$.
Unmatched silent steps decrease a well-founded measure.

\item \textbf{Final-state correspondence.}
If a state of $P_H$ is final, its related $P_L$ state
is final with the same result.
\end{enumerate}

Together, these lemmas establish a forward simulation. Writing $P\Downarrow B$
for observable behavior, $\mathsf{Safe}(B)$ for a non-wrong behavior, and
$\preceq$ for CompCert's behavior-improvement relation~\cite{compcertbehaviors}:
\[
\begin{aligned}
  &\mathcal{D}_{L\rightarrow H}(P_L)=\mathsf{OK}(P_H)
  \land P_H\Downarrow B_H 
  \rightarrow
  \exists B_L.\;
  P_L\Downarrow B_L
  \land B_H\preceq B_L.
\end{aligned}
\]

Unlike translation validation~\cite{pnueli1998translation,necula2000translation}, correctness is established
directly for the reconstruction algorithm rather than
by checking each candidate against an existing
compilation theorem.

\subsection{Translation Validation in Reverse}

Decompilation inexorably confronts the case where there are numerous plausible higher-level programs potentially consistent with a lower-level program (\emph{e}.\emph{g}., multiple looping constructs), potentially infinitely many (\emph{e}.\emph{g}., in the case of variable name identification). Traditional decompilers often employ tactics such as ranking candidates via heuristics and choosing the best supposed match. Instead, we employ translation validation to validate each proposed candidate. Translation validation is a well-known approach to constructing certified compilers by chaining together a set of \emph{validators} which establish that properties in the high-level fragment carry over to the low-level fragment~\cite{necula2000translation,tristan2008validators}. For example, CompCert's register allocator is implemented via translation validation: the register allocator need not be verified, but a certified validator checks the transformation from virtual registers to physical registers and stack locations~\cite{rideau2010regalloc}. 
Let $\mathcal{C}_{H\rightarrow L}$ be a CompCert compilation pass from a
higher-level language $H$ to a lower-level language $L$. Given a supplied
lower-level program $P_L$ and a reconstructed candidate $P_H$, the validator checks $\mathsf{validate}(P_L,P_H)=\mathsf{true}.$
Each validator establishes a pass-specific certified correspondence
$\mathcal{V}(P_H,P_L):
  \mathsf{validate}(P_L,P_H)=\mathsf{true}
  \Longrightarrow
  \mathcal{V}(P_H,P_L).$
We then prove that this correspondence implies a simulation:
$\mathcal{V}(P_H,P_L)
  \Longrightarrow
  \mathsf{forward\_simulation}
    \bigl(\mathsf{Sem}_H(P_H),\mathsf{Sem}_L(P_L)\bigr).$

Once the typed \ccname{Asm} input is fixed, reconstruction and checking can be separated.
For a compiler pass $\mathcal C_{H\rightarrow L}$, an exact round-trip
validator establishes
\[
\begin{aligned}
  &\mathsf{validate}(P_L,P_H)=\mathsf{true}
  \land P_H\Downarrow B_H
  \rightarrow
  \exists B_L.\;
  P_L\Downarrow B_L
  \land B_H\preceq B_L.
\end{aligned}
\]

Exact equality is one certification mechanism. Register allocation instead
uses CompCert's allocation relation: a proved validator checks an unverified
allocator's transformation from pseudo-registers to physical registers and
stack locations, including spills and reloads~\cite{rideau2010regalloc}.
Source validators compare modulo specific proved normalizations. Directly
verified passes construct a state relation and prove initial-state, step, and
final-state correspondence. Silent
administrative steps require a decreasing measure to rule out unmatched
infinite stuttering. These mechanisms have different local proofs but must
supply a continuous simulation path for the actual connected output.

\begin{figure*}[t]
    \centering
    \includegraphics[width=\textwidth]{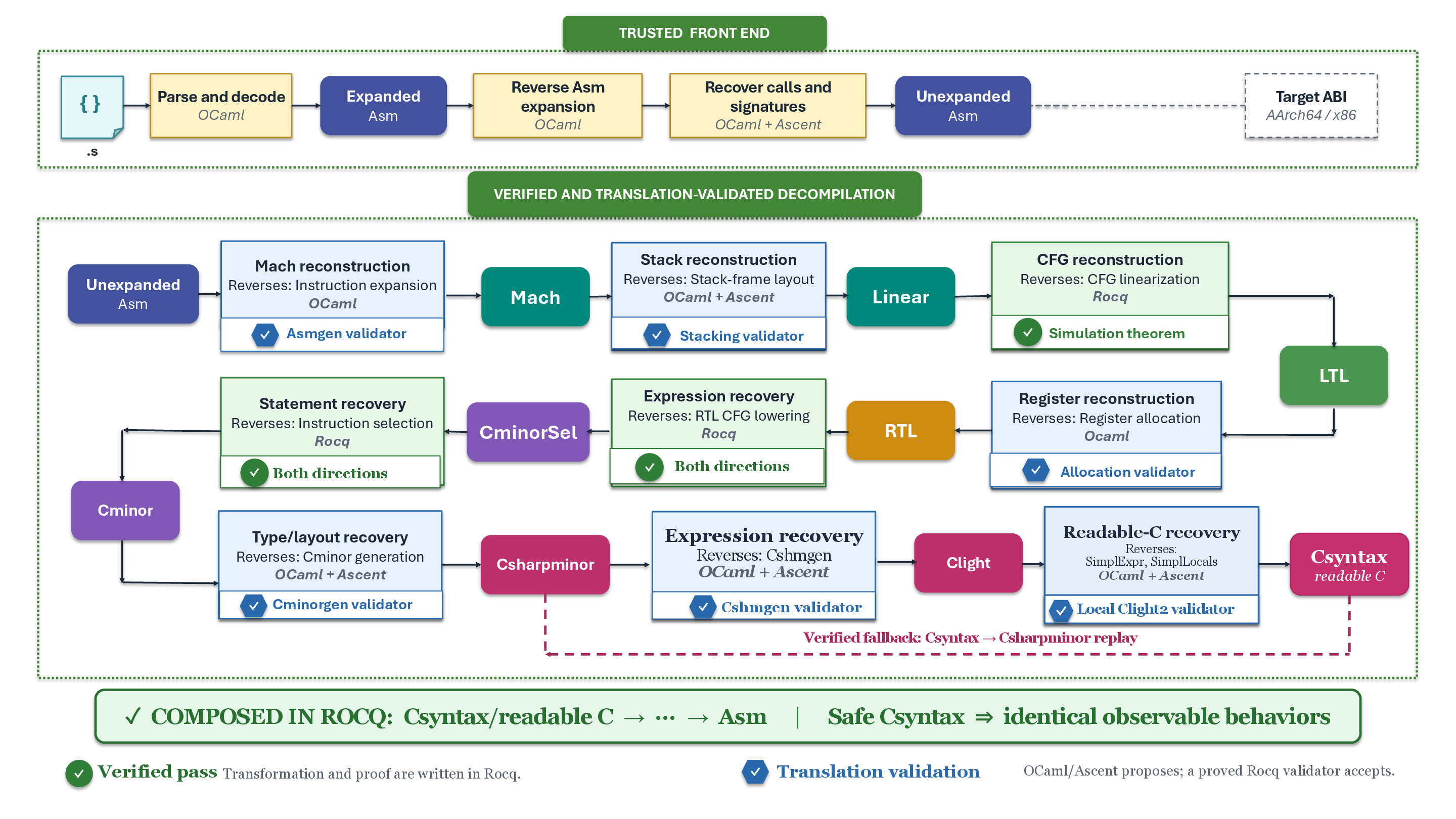}
    \caption{\tool{}'s high-level architecture and trusted computing base.}
    \Description{A trusted assembly-text front end constructs typed unexpanded Asm. Nine reverse transformations reconstruct Csyntax using verified passes and proved validators, with a source-validation fallback through Csharpminor.}
    \label{fig:compcert_decompilation}
\end{figure*}

\section{Decompiling CompCert}\label{sec:decompilation}



\tool{}'s high-level architecture is sketched in Figure~\ref{fig:compcert_decompilation}. Our implementation begins with a trusted front end that parses CompCert-emitted assembly text, reverses instruction expansion, and constructs CompCert's typed, unexpanded \ccname{Asm} representation. Precisely identifying reachable instructions in arbitrary binaries is undecidable in general~\cite{engel2024decidability}; we treat certified binary lifting~\cite{verbeek2022lifting,verbeek2024lifting} as orthogonal.
Function and call signatures are inferred and ranked by an application-wide
analysis of the assembly and target ABI. Checked platform and linked-binary proposals can augment this
ranking without entering the trusted core. In every configuration, the
resulting \ccname{Asm} program enters the same nine
semantic reverse transformations shown in
Figure~\ref{fig:compcert_decompilation}; the reconstruction algorithms,
validators, and semantic theorems after that boundary are unchanged.

Whenever the composed reconstruction function returns a \ccname{Csyntax} program,
\tool{} establishes an
unconditional guarantee that each reconstructed-C behavior is matched by a
behavior of the supplied lower-level program that is at least as defined. If the reconstructed
\ccname{Csyntax} program is safe, Section~\ref{sec:formalization} strengthens this
guarantee to observational equivalence. Reconstruction does not attempt to recover the exact intermediate
program that originally produced the input. For example, freshly generated
\ccname{RTL} pseudo-registers during decompilation from \ccname{LTL} to \ccname{RTL} need not coincide
with those of the original \ccname{RTL} program.


\subsection{\ccname{Asm} to \ccname{Mach}}

\ccname{Asm} is CompCert's representation of target assembly, AArch64 in our case.
Our input is its \emph{unexpanded} form, which still includes
pseudo-instructions for actions such as allocating and releasing a stack frame.
\ccname{Mach} is the intermediate language immediately before assembly
generation~\cite{10.1145/1538788.1538814}. Each function is a linear sequence
of instructions over machine registers, with explicit labels, branches, and
concrete stack offsets. \ccname{Mach} retains a more abstract view of operations,
calls, and returns: a single \ccname{Mach} instruction can correspond to several \ccname{Asm}
instructions.

%
Our \ccname{Asm}-to-\ccname{Mach} decompiler reconstructs candidates by recognizing the
instruction sequences emitted by $\mathsf{Asmgen}$. For
example, one \ccname{Mach} conditional branch can become an \ccname{Asm} comparison that sets
condition flags, followed by a branch that tests those flags. Because an assembly pattern
may admit several \ccname{Mach} reconstructions, we propose multiple candidates
for subsequent validation.
Our \ccname{Asm}-to-\ccname{Mach} validator reruns CompCert's \ccname{Asmgen} pass on the
reconstructed \ccname{Mach} program $P_M$. The validator accepts only if the result is exactly the
supplied unexpanded \ccname{Asm} program $P_A$:
\[
  \mathsf{validate}(P_A,P_M)=\mathsf{true}
  \Longrightarrow
  \mathsf{Asmgen}(P_M)=\mathsf{OK}(P_A).
\]

Validator soundness follows by unfolding the executable check and using
decidable equality of complete \ccname{Asm} programs. We then apply CompCert's existing
$\mathsf{Asmgen}$ correctness theorem, which gives a forward simulation from
the semantics of $P_M$ to the semantics of $P_A$.
Appendix~\ref{sec:appendix} gives the local semantic-preservation statements and proofs.

\subsection{Reconstructing Stack Frames: \ccname{Mach} to \ccname{Linear}}
\label{sec:mach-linear}

The \ccname{Mach}-to-\ccname{Linear} pass reverses CompCert's stack layout
pass, which transforms \ccname{Linear} into \ccname{Mach} by assigning concrete stack-frame offsets to abstract stack locations and inserting instructions to save and restore callee-saved registers. It also adjusts stack-relative operations and determines the final stack-frame size. In \ccname{Linear}, stack locations are represented abstractly
as $\mathsf{S(sl,ofs,\tau)}$, where
$\mathsf{sl}\in\{Incoming,Outgoing,Local\}$ identifies the
stack region, $\mathsf{ofs}$ is the offset, and $\tau$ is
the value type.
In the \texttt{process} example (Figure~\ref{fig:process-example}),
\ccname{Stacking} saves the caller's value of \textsf{x19}
at stack offset 16 and restores it before both exits: the tail call to
\(\mathsf{finish}\) and the return. These three stack operations are introduced
by \ccname{Stacking} and have no counterparts in the \ccname{Linear} program.

\paragraph{Callee-save facts and rules.}
To help us infer the stack layout, we use a logic program written in Datalog~\cite{10.1145/3622840}; our Datalog program implements the rules in Figure~\ref{fig:save_restore_rules}. The rules make use of facts supplied by an OCaml-based extractor, which traverses the \ccname{Mach} instruction and transforms them into facts that record stack stores (\textsf{setStack}) and loads (\textsf{getStack}), returns, tail calls, instruction positions, and
callee-saved registers. In the rules, $i$ is an instruction position, $r$ a machine register, $o$ a stack offset, and $\tau$ a memory type The instruction order is encoded via \textsf{next}, and returns and tail calls are treated uniformly as exits.
The rules run to a fixed point to compute
the contiguous save prefix for each function entry and the restore suffix before
each exit; the analysis pairs a save and restore only when their register, offset, and
memory type agree. The resulting matches guide reconstruction of the \textsf{Linear} program, whose correctness is independently verified in Rocq.
\begin{figure}[b]
\footnotesize
\[
\begin{array}{@{}c@{}}
\xinferrule*[right=\textsc{S-Base}]
  {\mathsf{firstInstr}(i) \quad
   \mathsf{setStack}(i,r,o,\tau)}
  {\mathsf{savePrefix}(i)}
\quad
\xinferrule*[right=\textsc{S-Step}]
  {\mathsf{savePrefix}(i) \quad
   \mathsf{next}(i,j) \quad
   \mathsf{setStack}(j,r,o,\tau)}
  {\mathsf{savePrefix}(j)}
\\[2ex]
\xinferrule*[right=\textsc{R-Base}]
  {\mathsf{getStack}(i,o,\tau,r) \quad
   \mathsf{next}(i,e) \quad
   \mathsf{exitAt}(e)}
  {\mathsf{restoreSuffix}(i,e)}
\quad
\xinferrule*[right=\textsc{R-Step}]
  {\mathsf{getStack}(i,o,\tau,r) \quad
   \mathsf{next}(i,j) \quad
   \mathsf{restoreSuffix}(j,e)}
  {\mathsf{restoreSuffix}(i,e)}
\\[2ex]
\xinferrule*[right=\textsc{Save-Restore}]
  {\mathsf{setStack}(i_s,r,o,\tau) \quad
   \mathsf{savePrefix}(i_s) \quad
   \mathsf{getStack}(i_r,o,\tau,r) \\
   \mathsf{restoreSuffix}(i_r,e) \quad
   \mathsf{calleeSaved}(r) \quad
   i_s < i_r}
  {\mathsf{saveRestore}(i_s,i_r,r,o,\tau)}
\end{array}
\]
\caption{Relational rules for identifying candidate save--restore pairs.}
\label{fig:save_restore_rules}
\end{figure}

\textit{Concrete-to-abstract stack offsets.}
If the local region
begins at byte offset $88$, for example,\\
$\mathsf{Mgetstack\ 96\ Tint\ X1}$ reconstructs the \ccname{Linear} local slot
$(96-88)/4=2$. An access in the fixed-frame area is rejected because it has no
\ccname{Linear} stack-location counterpart. \tool{} obtains these boundaries from a
\ccname{Linear} skeleton that temporarily omits unresolved stack accesses, then makes a
second pass to reconstruct them. Operations and addressing modes can also refer to \ccname{Linear}'s addressable stack
data. If that region begins at $\mathsf{fe\_stack\_data}$, the corresponding
\ccname{Linear}-relative offset is
$
  o_{\mathsf{Linear}} =
  o_{\mathsf{Mach}} - \mathsf{fe\_stack\_data}.
$
The implementation then applies CompCert's existing stack-shift functions to
operations and addressing modes with the negated displacement.

\paragraph*{Recovering frame sizes}

Recovering the \ccname{Linear} stack-data size requires a separate search. \ccname{Linear}
records only the size of its addressable stack-data area, whereas \ccname{Mach} records
the size of the complete activation record. \ccname{Stacking} also aligns the complete
frame, so several nearby \ccname{Linear} sizes can produce the same \ccname{Mach} frame size.
For a supplied \ccname{Mach} function $f_M$, \tool{}
tries candidate sizes in decreasing order. For each size $s$, it constructs a
\ccname{Linear} function $f_L$ with $\mathsf{fn\_stacksize}=s$, recomputes the frame layout,
reverses the corresponding stack-offset adjustments, and invokes the \ccname{Stacking}
validator. The first candidate for which
$\mathsf{Stacking}(f_L)=\mathsf{OK}(f_M)$
is accepted.
Before returning a reconstructed program, we perform the same validation at
the program level. For an input \ccname{Mach} program $P_M$ and reconstructed
\ccname{Linear} program $P_L$, we prove that the executable validator exactly
reflects this condition:
$
\mathsf{validate}(P_M,P_L)=\mathsf{true}
  \Longleftrightarrow
  \mathsf{Stacking}(P_L)=\mathsf{OK}(P_M).
$
This check covers not only reconstructed function bodies but also global
variables, external declarations, public symbols, their ordering, and the main
symbol. CompCert's correctness theorem for $\mathsf{Stacking}$ then provides a
forward simulation from the semantics of $P_L$ to the semantics of $P_M$.

\subsection{Reconstructing Control Flow: \ccname{Linear} to \ccname{LTL}}
\label{sec:linear-ltl}

The \ccname{Linear}-to-\ccname{LTL} decompiler reverses CompCert's
linearization pass, which transforms an \ccname{LTL} control-flow
graph into a sequential list of \ccname{Linear} instructions.
In \ccname{LTL}, each node contains a basic block with explicit
successors, whereas \ccname{Linear} uses labels, jumps, and
implicit fall-through.

Our decompiler reconstructs the \ccname{LTL} control-flow graph
in two passes. First, it assigns a fresh \ccname{LTL} node to
each \ccname{Linear} instruction and constructs a mapping from
labels to their corresponding nodes. Since a \ccname{Linear}
label is a pseudo-instruction, a label maps to the node
immediately following it, rather than to the node
assigned to the label itself.

In the second pass, the decompiler reconstructs each
\ccname{Linear} instruction as an \ccname{LTL} basic block with explicit
successors. Ordinary instructions receive the next
node as their successor. Conditional branches receive
two successors: the target obtained from the label
mapping and the fall-through node. Unconditional jumps
become \ccname{LTL} branches to the corresponding target nodes.
Jump tables are reconstructed by replacing their
target labels with graph nodes, while returns and
tail calls terminate execution without successors.
Label pseudo-instructions become branch-only blocks
leading to the following node. The algorithm also introduces a final sentinel node
representing the position immediately after the \ccname{Linear}
instruction sequence. Missing or duplicate labels
cause reconstruction to fail, ensuring that each
reconstructed branch has a uniquely determined target.

Unlike the translation-validated passes, the
\ccname{Linear}-to-\ccname{LTL} reconstruction algorithm and its
correctness proof are implemented entirely in Rocq.
The proof relates each reconstructed \ccname{LTL} node to
the suffix of the original \ccname{Linear} instruction list
beginning at the corresponding position. Corresponding call and return states
have the same register values and memory.

The main step proof proceeds by induction over the \ccname{LTL} transition.
When the transition executes an actual instruction, the correspondence between
\ccname{LTL} nodes and \ccname{Linear} code positions identifies the matching
\ccname{Linear} instruction. Executing that instruction produces the same trace
and re-establishes the state relation. Some \ccname{LTL} transitions, such as
entering a reconstructed block or traversing an inserted branch-only block,
do not advance the \ccname{Linear} program. We assign these states a
natural-number rank and prove in Rocq that every such silent step strictly
decreases the rank. Initial states are related, and related final states
return the same result. CompCert's \texttt{forward\_simulation\_star} theorem~\cite{compcertsmallstep}
combines these properties into a forward simulation.

\begin{theorem}[\ccname{Linear}-to-\ccname{LTL} semantic preservation]
Let $P_L$ be a \ccname{Linear} program and let
$\mathcal{D}_{\mathsf{Linear\rightarrow LTL}}(P_L)
  =\mathsf{OK}(P_T)
$.
Then
$
  P_T\Downarrow B_T
  \rightarrow
  \exists B_L.\;
  P_L\Downarrow B_L
  \land B_T\preceq B_L.
$
\end{theorem}

%
\subsection{Reconstructing Register Allocation: \ccname{LTL} to \ccname{RTL}}
Register allocation replaces \ccname{RTL} pseudo-registers with
physical registers and stack slots, introducing additional
moves, spills, and reloads. Reversing this transformation
is challenging because the original pseudo-registers are
lost, and the same physical location may hold different
values across control-flow paths. We follow a
translation-validation approach: an untrusted OCaml
decompiler generates \ccname{RTL} candidates, which are checked
using CompCert's formally verified register-allocation
validator~\cite{rideau2010regalloc}. Successful validation guarantees a forward
simulation from the reconstructed \ccname{RTL} to the original \ccname{LTL}.

For each basic block, reconstruction carries an immutable environment that
maps \ccname{LTL} locations to \ccname{RTL} pseudo-registers. A move introduced by allocation
does not become an \ccname{RTL} computation: it returns a new environment in which the
source and destination locations name the same pseudo-register. A real
operation reads its operands through the current environment and binds its
result location to a fresh pseudo-register. Locations first observed at a
block entry receive placeholder registers for live-in values. After the
control-flow graph has been explored, the output environments of predecessors
are reconciled at joins and loop headers; pairs introduced for
\texttt{Twolong} values are tracked together during this reconciliation.
Finally, \ccname{RTL} typing and \texttt{Allocation.check\_function} check the complete
candidate against the supplied \ccname{LTL} function.

%
\subsection{Reconstructing High-Level Program Structure}
\label{sec:high-level-reconstruction}
The remaining stages of \tool{} reconstruct higher-level
program abstractions from \ccname{RTL} through \ccname{Csyntax}. These passes recover expression
trees, generic operations,
addressable local variables, types, and source-level control-flow patterns when
they are present in the supplied intermediate representation. \ccname{RTL}-to-\ccname{CminorSel} and
\ccname{CminorSel}-to-\ccname{Cminor} use deductive verification. 

\paragraph{\textbf{Control-flow expressions.}}
\ccname{RTL} represents a function as a control-flow graph in which each node contains one operation, load, store, call, branch, builtin, jump table, or return. \ccname{CminorSel} instead combines expressions into statements. The verified \ccname{RTL}-to-\ccname{CminorSel} pass reconstructs an expression or statement for each \ccname{RTL}
node and uses labels and gotos to preserve joins, back edges, and explicit successors. Loads and stores retain their memory chunks and addressing modes; calls retain direct or indirect targets and argument order; jump tables become
\ccname{CminorSel} switch statements with the same indexed successors.

An \ccname{RTL} transition can correspond to several \ccname{CminorSel} transitions. For example, the \ccname{RTL} node \texttt{r3 = r1 + r2} is reconstructed as an assignment
whose operands are \texttt{Evar r1} and \texttt{Evar r2}, followed by a goto to the successor label. The assignment performs the \ccname{RTL} operation, while
sequencing, label lookup, and the goto introduce silent administrative steps.
The direct simulation matches one \ccname{RTL} transition with one or more \ccname{CminorSel} transitions.
%
Whole-pipeline correctness also requires a simulation from the reconstructed \ccname{CminorSel} program to \ccname{RTL}. We establish this using a phase-indexed state relation. Each reconstructed operation is matched by its corresponding \ccname{RTL} instruction, while intermediate execution steps may stutter only when a well-founded index decreases. A register-environment invariant ensures that used \ccname{RTL} registers agree with their corresponding \ccname{CminorSel} locals. Evaluation-correspondence lemmas cover operations, addresses, conditions, calls, memory effects, builtins, returns, and jump-table destinations. The simulations in both directions are presented in Appendix~\ref{sec:appendix}.

\paragraph{\textbf{Instruction selection.}}
This pass reconstructs \ccname{Cminor} from \ccname{CminorSel}. \ccname{CminorSel} uses target-specific
operations and addressing modes, whereas \ccname{Cminor} represents
computations using generic operators and explicit address
expressions. The decompiler recursively translates \ccname{CminorSel}
expressions and statements into \ccname{Cminor}.
Composite selected operations may require multiple \ccname{Cminor} operations.
A key challenge arises from expression-level
constructs such as let bindings, conditional
expressions, and calls, which cannot always be
represented by a single \ccname{Cminor} expression. The
decompiler introduces fresh temporary variables
and statements to preserve their evaluation order.
For example, the \ccname{CminorSel} expression
$\mathsf{let}\ t=x+3\ \mathsf{in}\ t*t$
is reconstructed as:
\texttt{tmp = x + 3; result = tmp * tmp;}.
The temporary ensures that the bound expression
is evaluated only once. Fresh identifiers are
chosen above all identifiers occurring in the
input function to avoid conflicts with existing
variables.

The reconstruction algorithm and its correctness
proof are implemented in Rocq. The proof relates
the original \ccname{CminorSel} environment to the
reconstructed \ccname{Cminor} environment, allowing
additional temporary variables while preserving
the values of existing variables. It also relates
expression-level let bindings to their corresponding
generated temporaries. These relations are used
to prove that reconstructed expressions and
statements preserve evaluation order, computations,
memory operations, and observable events. The proof establishes a forward simulation from
the supplied \ccname{CminorSel} program to the reconstructed
\ccname{Cminor} program, under explicit premises requiring
defined unsigned comparisons, defined selections,
and unique labels. These premises belong to this
pass-local simulation direction. The complementary
\ccname{Cminor}-to-\ccname{CminorSel} simulation used for the connected theorem
derives the required facts from each successful \ccname{Cminor} step
and has no global definedness premise.

\paragraph{\textbf{Recovering addressable variables.}}
The \ccname{Cminor}-to-\ccname{Csharpminor} decompiler reconstructs
each function's nonempty stack block as a single addressable variable.
CompCert's \ccname{Cminorgen} pass allocates \ccname{Csharpminor}
addressable locals, each with an identifier and size, at offsets within
a single \ccname{Cminor} stack block. This erases the original variable
boundaries, which therefore cannot be uniquely recovered.
The reconstructed variable therefore has the size of the \ccname{Cminor}
stack block and a fresh identifier chosen above all global, parameter,
and local-variable identifiers to avoid conflicts. Functions without
stack space require no additional addressable variable. For a nonempty
stack block, offset zero becomes the variable's address, and each interior
address becomes a constant byte displacement from that base.

%
The validator checks this reconstruction by recompiling the untrusted
generator's \ccname{Csharpminor} candidate $P_H$ with \ccname{Cminorgen}.
It requires at most one nonnegative-size addressable object per function
and accepts when the generated \ccname{Cminor} program $P_G$ equals the
supplied program $P_L$, either directly or after normalizing constant
stack-address displacements with $N_s$:
$$
  \mathsf{Cminorgen}(P_H)=\mathsf{OK}(P_G)
  \land
  \bigl(P_G=P_L\lor N_s(P_G)=N_s(P_L)\bigr).
$$
We prove simulations in both directions for $N_s$ and compose them with
\ccname{Cminorgen} correctness; Appendix~\ref{sec:appendix} gives the
validator and proof. Successful validation therefore establishes a
forward simulation from the reconstructed \ccname{Csharpminor} program
to the supplied \ccname{Cminor} program.

The single-object check is a conservative output-quality and safety
guard. The current forward-simulation proof does not use it:
\ccname{Cminorgen} correctness and the proved $N_s$ simulations suffice.
We retain the check to avoid extra object boundaries that can introduce
additional source-level undefined behavior and make the safety premise
for observational equivalence harder to establish.

\paragraph{\textbf{Recovering types and structured statements.}}
The \ccname{Csharpminor}-to-\ccname{Clight} decompiler reverses \ccname{Cshmgen}. It must
recover types erased by lowering and recognize the blocks, loops, and numbered
exits used to encode structured control flow. An OCaml extractor records type
facts from signatures, operations, memory accesses, and assignments;
Ascent propagates these facts across related variables. The inferred facts
guide candidate declarations and expressions.
Addressable objects retain their byte size but not their source element type.
Access widths and scaled addresses therefore guide array candidates, with a
byte array used when the access patterns do not suggest a more precise layout.
Recognized \texttt{block}/\texttt{loop}/\texttt{exit} patterns can become \ccname{Clight}
conditionals, loops, breaks, and switches. However, the connected
\ccname{RTL}-to-\ccname{CminorSel} lifter emits one labeled statement per CFG node
and connects nodes with gotos. It therefore does not supply these structured
patterns to the later recognizer. The evaluated end-to-end pipeline retains
label/goto control flow and recovers no structured loops or source-level
switches; the recognizer applies only when such patterns are present in a
separately supplied candidate.

The difficult part is that exact round-trip equality is too restrictive.
Discarded call destinations and the representation of narrow ABI values can
make the generated \ccname{Csharpminor} program differ syntactically from the supplied
one, $P_S$. The validator therefore recompiles the candidate $P_C$ with \ccname{Cshmgen}
and accepts only when the complete result belongs to a finite family
$\mathcal N(P_S)$ obtained using proved discarded-result, narrow-value, and
parameter-entry transformations. Each transformation has a reverse
simulation, so accepting these additional forms does not weaken soundness.
A separate arity check is necessary because \ccname{Cshmgen} retains actual
variadic arguments but can erase whether a prototype declared the correct
fixed prefix.

The OCaml and Ascent analyses remain untrusted. Successful validation implies
\[
\begin{aligned}
\mathsf{validate}(P_S,P_C)=\mathsf{true}
\Longrightarrow
\mathsf{forward\_simulation}
\bigl(\mathsf{Sem}_{\mathsf{Clight}}(P_C),
      \mathsf{Sem}_{\mathsf{Csharpminor}}(P_S)\bigr).
\end{aligned}
\]
Thus, wrong inferred types or structures can cause rejection but cannot make
an accepted reconstruction unsound. Appendix~\ref{sec:appendix} defines the
accepted normalizations and gives the proof.

\paragraph{\textbf{Recovering recompilable C}}
The accepted \ccname{Clight} reconstruction $P_L$ guides the final recovery of a \ccname{Csyntax}
program $P_C$, including source-level expression grouping, loop initializers,
short-circuit expressions, and comma expressions. The generator is untrusted,
so the final validator recompiles the candidate through \ccname{SimplExpr} and
\ccname{SimplLocals}. It accepts when these verified passes reproduce the
complete accepted \ccname{Clight} program exactly, or when a fixed family of proved
normalizations makes the generated program equal to it. These remove
administrative \texttt{skip} prefixes and guarded call-result copies, and fold
value-preserving assignment casts under precise constant/range checks. Their
simulations are proved separately. Writing
$\mathsf{validate}_{\mathsf{Clight\rightarrow C}}$ for this local check:
\[
\begin{aligned}
\mathsf{validate}_{\mathsf{Clight\rightarrow C}}(P_L,P_C)
  =\mathsf{true} \Longrightarrow
\mathsf{forward\_simulation}
\bigl(\mathsf{Sem}_{\mathsf C}(P_C),
      \mathsf{Sem}_{\mathsf{Clight}}(P_L)\bigr).
\end{aligned}
\]
On the preferred source path, the \ccname{Csharpminor}-to-\ccname{Clight} validator uses the
same $P_L$, so this theorem closes the final adjacent boundary. If the local
comparison rejects, a proved fallback instead replays the \ccname{Csyntax} candidate
through \ccname{SimplExpr} and \ccname{SimplLocals}, applies only proved
cleanup combinations, and validates the result against the reconstructed
\ccname{Csharpminor} input. The complete composition accepts either source-side proof
premise.

\subsection{Recovering Function Signatures from Assembly}
\label{sec:signature-recovery}

CompCert's typed \ccname{Asm} requires signatures for functions and calls,
even when the input assembly carries no type information. \tool{} recovers
these interfaces by reasoning from about functions their callers.
Assembly does not determine a unique signature: a 64-bit value may be
an integer or a pointer, and unused results leave the return type unconstrained.
The analysis therefore ranks alternatives for reconstruction to try to validate.
We first use the target calling convention to identify potential arguments
and results. Reading an argument register before its first write suggests a
parameter, and using a call's result register before overwriting it suggests a
returned value.
A Datalog-based analysis propagates these analysis facts across direct
calls throughout the application, connecting each callee's interface to
observations at its call sites.
When a function forwards a call's result, the analysis also connects their
return types. These rules iterate until no new analysis information is found. 

Where choices remain, platform declarations for imported functions~\cite{applearm64,arm2026aapcs64} and signatures proposed by Ghidra~\cite{ghidra} or IDA/Hex-Rays~\cite{idapro} guide ranking. \tool{} restricts platform declarations to actual imports and checks decompiler proposals against independently decoded function entries and direct calls. 
Each search candidate assigns signatures to functions and call sites together,
producing a complete typed \ccname{Asm} program. The search starts from a default
assignment, then tries individual changes and combinations in rank order, up
to a fixed candidate budget.
When validation rejects a function, the search prioritizes changes to related
signature choices while leaving unrelated choices fixed. A successful assignment
supports reconstruction through the same nine reverse passes. 
Acceptance establishes the typed-\ccname{Asm}-to-C guarantee, but does not
establish that the original signatures have been recovered. An incorrect
interface can still be internally consistent enough to pass validation.
The theorem therefore relates reconstructed C to the selected typed assembly
candidate; ensuring that this candidate faithfully represents the input
assembly remains an obligation of the trusted front end
(Section~\ref{sec:formalization}).

\section{Establishing End-to-End Soundness}
\label{sec:formalization}

The mechanized theorem begins with a typed, unexpanded CompCert \ccname{Asm} program and
ends with reconstructed \ccname{Csyntax}. Parsing textual assembly, reversing
CompCert's instruction expansion, assigning function and call signatures,
decoding binaries, linking, and relocation precede this boundary. Within it, eight fixed
lower/\ccname{Clight} checks establish simulations between adjacent CompCert semantics.
At the recompilable-C boundary, the proof uses either the adjacent local \ccname{Clight2}
check or a verified replay from \ccname{Csyntax} directly to the retained \ccname{Csharpminor}
program.

The word \emph{typed} describes the \ccname{Asm} value consumed by the semantics; it
does not require those types to be retained from compilation. This gives two
layers. The trusted front end parses textual assembly and constructs the typed
\ccname{Asm} program, including its calling interfaces. The certified core then relates
that typed \ccname{Asm} program to the reconstructed C program. \tool{} mechanizes the
second layer; proving the parser is outside the scope of this work. In the
signature-free mode of Section~\ref{sec:signature-recovery}, candidate
interfaces are inferred by application-wide assembly analysis, optionally
augmented with checked platform and linked-binary proposals, before this
formal boundary. They are then examined through validation and native
experiments.

The theorem below is universally quantified over $P_{\mathsf{Asm}}$, so the
same typed-\ccname{Asm}-to-C proof applies whether its signatures were retained or
inferred. This quantification must not be read as a proof of the first
front-end layer. An inferred candidate can be internally consistent and still
misrepresent an interface that is unobservable in the text. In that case the
theorem remains about the candidate typed \ccname{Asm} semantics, rather than the
external executable semantics.

\subsection{Composing Local Guarantees}

Let $\mathcal L_0=\mathsf{Asm},\ldots,
\mathcal L_8=\mathsf{Clight}$ follow the first eight reconstruction stages, and let
\[
  \mathcal D_i:\mathcal L_i\rightarrow\mathsf{res}(\mathcal L_{i+1})
  \qquad(0\leq i<8)
\]
denote their certified interfaces. For a translation-validated pass,
$\mathsf{OK}$ includes acceptance by the proved validator. The connected
driver retains both the reconstructed \ccname{Csharpminor} and \ccname{Clight} programs for the
source-boundary procedure. That procedure returns \ccname{Csyntax} only when either the
preferred local \ccname{Clight2} validator accepts or the verified
\ccname{Csyntax}-to-\ccname{Csharpminor} replay accepts. We write $\mathcal D_{\mathsf{pipe}}$
for the resulting error-propagating composition. Thus
$\mathcal D_{\mathsf{pipe}}(P_{\mathsf{Asm}})=
\mathsf{OK}(P_{\mathsf C})$ states that all eight fixed checks accepted, one
proved source-boundary mode accepted, and every stage consumed the actual
output retained from its predecessor.

Composition requires a continuous semantic path from \ccname{Csyntax} to \ccname{Asm}. \tool{}
establishes both relevant simulation directions at the \ccname{RTL}, \ccname{CminorSel}, and
\ccname{Cminor} boundaries. The decompilation-oriented theorems express the local
relationship between the input and reconstruction. The complementary
compiler-oriented simulations connect \ccname{Cminor} through \ccname{CminorSel} to \ccname{RTL} and
therefore align with the surrounding CompCert results.

The complementary \ccname{Cminor}-to-\ccname{CminorSel} simulation is the most technically
demanding new proof in \tool{}. It cannot be obtained by instantiating
CompCert's instruction-selection theorem: the supplied \ccname{CminorSel} program is
not required to be the exact result of applying instruction selection to the
reconstructed \ccname{Cminor} program. Reconstruction can introduce different
temporary assignments, sequencing, labels, and administrative control flow.
We therefore relate their executions directly. A phase-indexed relation
records whether execution is evaluating setup assignments, performing the
corresponding action, or following an administrative label or goto. Action
phases match the \ccname{RTL} instruction from which the code was reconstructed;
administrative phases take silent steps and decrease a well-founded measure.
\ccname{RTL} provenance rules out effectful expression forms that the pass never
produces. The same proof structure covers calls, memory operations,
conditions, builtins, and jump tables without adding a global definedness
premise. Appendix~\ref{app_sec:formal-proofs} gives the detailed construction.

This modular proof does not require C output to recompile into assembly
that is syntactically identical to the input. Such a comparison would reject
differences in temporary names, labels, register choices, and stack layout
even when the programs have the required semantic relation. Instead, each
boundary checks the information available at that representation and reuses a
CompCert theorem. Some new
complementary simulations bridge the remaining boundaries. For SQLite, the reconstructed C program compiles successfully, but the local validator rejects it because the generated Clight2 program differs from the expected one in seven narrow assignments. These differences are handled by a proved normalization at the Csharpminor level, allowing the alternative validator to accept the reconstruction.

\begin{theorem}[Composed decompiler simulation]
\label{thm:composed-decompiler-simulation}
For every supplied unexpanded \ccname{Asm} program $P_{\mathsf{Asm}}$ and reconstructed
\ccname{Csyntax} program $P_{\mathsf C}$,
\[
\mathcal D_{\mathsf{pipe}}(P_{\mathsf{Asm}})=\mathsf{OK}(P_{\mathsf C})
\Longrightarrow
\mathsf{forward\_simulation}
  \bigl(\mathsf{Sem}_{\mathsf C}(P_{\mathsf C}),
        \mathsf{Sem}_{\mathsf{Asm}}(P_{\mathsf{Asm}})\bigr).
\]
Consequently, if $P_{\mathsf C}\Downarrow B_C$, then the supplied \ccname{Asm} program
has a behavior $B_A$ such that $B_C\preceq B_A$ under CompCert's
behavior-improvement relation.
\end{theorem}

Here and throughout the end-to-end statements,
$\mathsf{Sem}_{\mathsf C}(P)$ denotes
\texttt{Cstrategy.semantics} $P$, CompCert's strategy semantics for C. It does
not denote the nondeterministic \texttt{Csem} semantics. Atomic strategy
semantics is used internally when deriving the converse simulation under the
safety premise below.

The proof composes the semantic results for the actual
intermediate programs returned by
$\mathcal D_{\mathsf{pipe}}$; candidate-generation algorithms do not occur in
its conclusion. It also does not rely on testing, equality with the unavailable
original source, or recompilation into a second assembly program.
In particular, the conclusion has no premise requiring a retained source
signature. A candidate rejected by any checker receives no guarantee. An
accepted inferred candidate receives the guarantee relative to its typed \ccname{Asm}
semantics; establishing that this semantics is the faithful interpretation of
the supplied text is a responsibility of the trusted front end.

\subsection{Observational Equivalence}

The composed simulation gives an unconditional compiler-style guarantee for
every accepted reconstruction. For exact equivalence, CompCert's generic
results convert the simulation through atomic \ccname{Cstrategy} semantics because
\ccname{Cstrategy} semantics is receptive and \ccname{Asm} semantics is
determinate~\cite{compcertsmallstep}. As in verified
compilation, the C program must be safe: after source-level undefined behavior,
its semantics does not constrain the lower-level execution~\cite{compcertbehaviors}.

\paragraph{Why safety is required for exact equivalence.}
The composed simulation theorem does not require the reconstructed
C program to be safe. For every accepted reconstruction, it
guarantees that every defined observable behavior of the C program
is preserved by the supplied assembly. Safety is needed only for
the converse direction, which establishes exact observational
equivalence.

Exact round-trip validation alone does not establish safety.
For example, AArch64 hardware integer division returns zero when the
divisor is zero~\cite[SDIV, UDIV]{arm2024a64}, whereas division by zero is undefined in C~\cite{iso2011c}.
A reconstructed expression \texttt{x/y} may therefore recompile
exactly to the supplied division instruction, even though their
behaviors differ when \texttt{y = 0}. The round-trip check
establishes that the reconstruction compiles to the supplied
assembly, but CompCert's correctness theorem does not constrain
the target behavior once the source encounters undefined behavior.

This limitation is inherited from CompCert itself. Proving the
reconstructed C program safe rules out such undefined executions,
allowing us to establish exact observational equivalence.
Here, $\mathsf{Safe}(P)$ means that every behavior of $P$ is non-wrong.

\begin{theorem}[Whole-decompiler observational equivalence]
\label{thm:safe-composed-decompiler-simulation}
If the composed decompiler returns $P_{\mathsf C}$ and the reconstructed
program is safe, then it has exactly the observable behaviors of the supplied
\ccname{Asm} program:
\[
\mathcal D_{\mathsf{pipe}}(P_{\mathsf{Asm}})=\mathsf{OK}(P_{\mathsf C})
\land \mathsf{Safe}(P_{\mathsf C})
\Longrightarrow
\forall B.\;
\bigl(P_{\mathsf C}\Downarrow B
\Longleftrightarrow P_{\mathsf{Asm}}\Downarrow B\bigr).
\]
\end{theorem}

Safety is required only for the exact-equivalence corollary.
Without this assumption, Theorem~\ref{thm:composed-decompiler-simulation} still guarantees
that every defined observable behavior of the reconstructed
C program is preserved by the supplied typed \ccname{Asm} program.
However, it does not establish the converse inclusion,
and hence cannot transfer arbitrary universal behavioral
properties from C to \ccname{Asm}. We demonstrate the stronger result on a closed recovered
checksum program. Rocq verifies its complete finite C
execution graph, including allocations, loads, and stores
for five addressable locals, establishing safety and
the result 222. The exact-equivalence corollary then
establishes the same observable result for the supplied
\ccname{Asm} program.

\paragraph{Connecting the extracted checker to the theorem.}
The preceding results compose the local pass theorems, but an executable
decompiler must additionally ensure that one reported success supplies the
premises from one continuous candidate chain. In particular, the OCaml search
must not accidentally combine an accepted \ccname{Mach} candidate from one
attempt, a \ccname{Linear} candidate from another, and readable C from a third.
We therefore define and extract a single Rocq function
$\mathcal V_{\mathsf{conn}}$. It receives the supplied \ccname{Asm} program,
the proposed \ccname{Mach}, \ccname{Linear}, \ccname{Clight}, and
\ccname{Csyntax} programs, and a tag selecting one of the two proved
source-boundary modes. It reruns the \ccname{Asm}-to-\ccname{Mach} and
\ccname{Mach}-to-\ccname{Linear} validators, recomputes every intermediate
program from \ccname{LTL} through \ccname{Csharpminor}, and reruns the selected
readable-source check. The mode $\mathsf{Clight2}$ selects validation of the
reconstructed \ccname{Clight}-to-\ccname{Csyntax} boundary, while
$\mathsf{CsharpminorReplay}$ selects direct replay of the proposed
\ccname{Csyntax} program to \ccname{Csharpminor}. The checker returns the
proposed \ccname{Csyntax} program only when the selected connected check
succeeds.

Only the programs proposed at validation boundaries appear as arguments to
$\mathcal V_{\mathsf{conn}}$. The \ccname{LTL}, \ccname{RTL},
\ccname{CminorSel}, \ccname{Cminor}, and \ccname{Csharpminor} programs do not
appear as separate arguments because the checker recomputes them in sequence.
Thus, the smaller argument list still checks one continuous execution of all
nine reverse passes.

\begin{theorem}[Extracted connected-checker soundness]
\label{thm:connected-checker-soundness}
For supplied programs $P_{\mathsf{Asm}}$, $P_{\mathsf{Mach}}$,
$P_{\mathsf{Linear}}$, $P_{\mathsf{Clight}}$, and $P_{\mathsf C}$, source mode
$m\in\{\mathsf{Clight2},\mathsf{CsharpminorReplay}\}$, and program
$\widehat P_{\mathsf C}$ returned by the connected checker,
\[
\begin{split}
&\mathcal V_{\mathsf{conn}}
  (P_{\mathsf{Asm}},P_{\mathsf{Mach}},P_{\mathsf{Linear}},
   P_{\mathsf{Clight}},P_{\mathsf C},m)
   =\mathsf{OK}(\widehat P_{\mathsf C})\\
&\qquad\Longrightarrow\quad
  \widehat P_{\mathsf C}=P_{\mathsf C}\ \land\
  \mathsf{forward\_simulation}
  \bigl(\mathsf{Sem}_{\mathsf C}(\widehat P_{\mathsf C}),
        \mathsf{Sem}_{\mathsf{Asm}}(P_{\mathsf{Asm}})\bigr).
\end{split}
\]
\end{theorem}

The proof unfolds the extracted checker and considers each validation result.
A successful evaluation yields the two target-validation premises, the actual
outputs of all five recomputed intermediate transformations, and either the
local \ccname{Clight2} validation premises or the
\ccname{Csyntax}-to-\ccname{Csharpminor} replay premise. These are exactly the
premises of Theorem~\ref{thm:composed-decompiler-simulation}. Because every
successful branch of the connected checker returns
$\mathsf{OK}(P_{\mathsf C})$, the success equation also gives
$\widehat P_{\mathsf C}=P_{\mathsf C}$.

The OCaml driver uses this extracted function as its sole final acceptance
boundary and emits the returned program rather than independently interpreting
the individual validation results. It establishes that the concrete C program
emitted by the executable decompiler is exactly the program covered by the
composed semantic guarantee.

\paragraph{Trusted computing base.}
The trusted base of the certified core contains the Rocq kernel, the CompCert
semantics, and the assumptions inherited from CompCert. OCaml and Ascent
reconstruction procedures inside this boundary are untrusted because their
outputs are validated. For a claim beginning at an external \texttt{.s} file,
the textual assembly parser and the selected interface assignment are part of
the trusted front end. The printer, assembler, linker, and native compilers
also lie outside the theorem.
Appendix~\ref{sec:appendix} gives the local statements and state relations;
the supplement provides proof accounting, and the artifact contains the assumption audit.

\section{Implementation}
\label{sec:implementation}

We implement the decompiler as an extension of CompCert 3.17, reusing its
intermediate-language definitions, operational semantics, calling conventions,
forward transformations, and program printers. Each reverse pass consumes a
CompCert program and either returns a reconstructed higher-level program or
reports that reconstruction failed. The passes can be executed independently
or composed into an assembly-to-\ccname{Csyntax}/recompilable C pipeline.

Verified reconstruction passes are implemented and proved correct in Rocq.
For translation-validated passes, OCaml performs syntax-directed reconstruction,
while Ascent supports relational searches such as recovering stack
layouts, calls, and signatures. These components generate candidates but are
not trusted for correctness: a validator proved sound in Rocq checks every
accepted result. The reverse passes are integrated into CompCert's command-line driver through
separate decompilation flags. Intermediate printers expose the reconstructed
\ccname{Mach}, \ccname{Linear}, \ccname{LTL}, \ccname{RTL}, \ccname{CminorSel}, \ccname{Cminor}, \ccname{Csharpminor}, \ccname{Clight}, and
\ccname{Csyntax}/recompilable C
programs. The current implementation parses AArch64 assembly and uses the
target ABI when recovering machine-level information. When signatures are
unavailable, the driver parses components independently, namespaces private
symbols, constructs an application graph, and combines register-use,
stack-access, and call-site facts through an Ascent fixed point. It ranks complete
signature assignments and tries them under a cumulative candidate budget.
Versioned Apple and GNU/Linux profiles are filtered against checked image
imports, and independently decoded AArch64 entries and direct calls bind
Ghidra and IDA proposals to the linked program. These sources only add untrusted
candidate preferences. Each assignment produces a typed CompCert \ccname{Asm} candidate; no
source-derived signature is consulted during this search. Parsing and
candidate generation precede the mechanized semantic boundary. The
complete-pipeline proof starts from each candidate \ccname{Asm} value and covers the
nine reverse transformations through \ccname{Csyntax}, so the same proof applies to
every accepted signature-free result.

The implementation adds approximately 52,100 lines of Rocq code beyond CompCert 3.17, excluding tests. Of these, about 33,200 lines are dedicated to proofs, validators, and reverse simulations.
For the type/layout refinement evaluated in
Section~\ref{sec:type-layout-evaluation}, an analysis-only \ccname{Csharpminor} control-flow graph (CFG)
assigns separate identities to parameter entries and assignment, call, and
builtin results. It records a reaching identity only when all reachable
predecessors agree. Ascent propagates field-access and stride facts through
justified value copies, projecting it to declarations with one syntactic
definition; resource-budget skips retain the earlier proposals. Composite
declarations are shared only within flow-related families whose selected
layouts, including nested pointer targets, are compatible. Identical physical
fields alone do not establish a shared family or an original nominal type.
Candidate ranking first compares aggregate function costs lexicographically (casts, expression
nodes, and locals); only ties prefer fewer composite declarations and members.
All proposals still require acceptance by the unchanged \ccname{Cshmgen} and final
\ccname{Csyntax} validators.

\section{Evaluation}
\label{sec:evaluation}

We evaluate \tool{} through four research questions, \rq{1}--\rq{4}:

\begin{researchquestions}
\rqquestion{1}{How often does \tool{} reconstruct a complete chain from
assembly without retained signatures?}
\rqquestion{2}{How much do checked platform and decompiler proposals
improve coverage?}
\rqquestion{3}{Does the reverse pipeline succeed when exact signatures
are supplied?}
\rqquestion{4}{How does the pipeline generalize to unseen inputs, and how
does its C output behave and compare with the source?}
\end{researchquestions}

\subsection{\rq{1}: Assembly-Only Signature Inference}
\label{sec:no-signature-evaluation}

\iffalse
\begin{table*}[t]
\centering
\caption{Signature recovery and complete-chain coverage at \texttt{-O0}.}
\label{tab:no-signature-main}
\scriptsize
\setlength{\tabcolsep}{3pt}
\renewcommand{\arraystretch}{0.86}
\textbf{(a) First-signature agreement}\\[-2pt]
\begin{tabularx}{\textwidth}{@{}Xrr@{}}
\toprule
Application & \tool{} & Ghidra \\
\midrule
zlib 1.3.1 & 99/151 (65.6\%) & 81/151 (53.6\%) \\
Lua 5.4.7 & 600/1,079 (55.6\%) & 89/1,079 (8.2\%) \\
libuv 1.48.0 & 335/613 (54.6\%) & 116/613 (18.9\%) \\
libpng 1.6.43 & 237/530 (44.7\%) & 18/530 (3.4\%) \\
Binutils libiberty & 197/622 (31.7\%) & 165/622 (26.5\%) \\
BearSSL 0.6 & 252/605 (41.7\%) & 53/605 (8.8\%) \\
OpenSSL 3.5.9 (libssl) & 1,501/3,165 (47.4\%) & 392/3,165 (12.4\%) \\
OpenSSL 3.5.9 (libcrypto) & 5,440/14,674 (37.1\%) & 4,255/14,674 (29.0\%) \\
wolfSSL 5.9.4 & 679/1,322 (51.4\%) & 161/1,322 (12.2\%) \\
Mbed TLS 3.6.4 & 1,103/2,174 (50.7\%) & 196/2,174 (9.0\%) \\
nginx 1.30.5 & 601/1,407 (42.7\%) & 118/1,407 (8.4\%) \\
LibreSSL 4.3.3 & 2,718/6,716 (40.5\%) & 2,781/6,716 (41.4\%) \\
Coreutils 9.5 & 2,798/5,819 (48.1\%) & --- \\
curl 8.8.0 & 691/1,740 (39.7\%) & 129/1,740 (7.4\%) \\
BusyBox 1.36.1 & 647/2,131 (30.4\%) & 2,003/2,131 (94.0\%) \\
SQLite 3.46.0 & --- & 412/2,521 (16.3\%) \\
\midrule
Shared 14-application scope & 15,100/36,929 (40.9\%) & 10,557/36,929 (28.6\%) \\
\bottomrule
\end{tabularx}

\vspace{5pt}
\textbf{(b) Accepted complete chains}\\[-2pt]
\begin{tabularx}{\textwidth}{@{}Xrrrrr@{}}
\toprule
Application & \tool{} only & + Platform & + Ghidra & + IDA & Cumulative \\
\midrule
zlib 1.3.1 & 15/15 & 15/15 & 15/15 & 15/15 & 15/15 \\
Lua 5.4.7 & 26/34 & 26/34 & 26/34 & 28/34 & 28/34 \\
libuv 1.48.0 & 26/37 & 27/37 & 31/37 & 28/37 & 31/37 \\
libpng 1.6.43 & 16/16 & 16/16 & 16/16 & 16/16 & 16/16 \\
Binutils libiberty & 115/122 & 116/122 & 116/122 & 115/122 & 115/122 \\
BearSSL 0.6 & 263/277 & 264/277 & 264/277 & 264/277 & 264/277 \\
OpenSSL 3.5.9 (libssl) & 73/94 & 73/94 & 76/94 & 73/94 & 76/94 \\
OpenSSL 3.5.9 (libcrypto) & 525/980 & 579/980 & 630/980 & 637/980 & 777/980 \\
wolfSSL 5.9.4 & 33/42 & 33/42 & 34/42 & 33/42 & 34/42 \\
Mbed TLS 3.6.4 & 85/108 & 87/108 & 87/108 & 87/108 & 87/108 \\
nginx 1.30.5 & 67/120 & 82/120 & 85/120 & 97/120 & 97/120 \\
LibreSSL 4.3.3 & 313/480 & 337/480 & 374/480 & 387/480 & 389/480 \\
Coreutils 9.5 & 75/132 & 86/132 & 86/132 & 86/132 & 94/132 \\
curl 8.8.0 & 165/213 & 173/213 & 174/213 & 178/213 & 178/213 \\
BusyBox 1.36.1 & 77/493 & 304/493 & 311/493 & 363/493 & 405/493 \\
SQLite 3.46.0 & 0/1 & 0/1 & 0/1 & 0/1 & 0/1 \\
\midrule
All 16 applications & 1,874/3,164 (59.2\%) & 2,218/3,164 (70.1\%) & 2,325/3,164 (73.5\%) & 2,407/3,164 (76.1\%) & 2,606/3,164 (82.4\%) \\
\bottomrule
\end{tabularx}
\end{table*}
\else
\begin{table*}[t]
\centering
\caption{Signature recovery and complete-chain coverage at \texttt{-O0}.}
\label{tab:no-signature-main}
\small
\setlength{\tabcolsep}{1.6pt}
\renewcommand{\arraystretch}{1.0}
\begin{tabularx}{\textwidth}{@{}>{\raggedright\arraybackslash}Xrrr@{\hspace{8pt}}rrrrrr@{}}
\toprule
& \multicolumn{3}{c}{\textbf{(a) First-signature agreement}}
& \multicolumn{6}{c}{\textbf{(b) Accepted complete chains}} \\
\cmidrule(lr){2-4}\cmidrule(l){5-10}
Application & $N$ & \tool{} & Ghidra
& $N$ & Only & + Plat. & + Ghidra & + IDA & Total \\
\midrule
zlib 1.3.1 & 151 & 99 (65.6\%) & 81 (53.6\%) & 15 & 15 & 15 & 15 & 15 & 15 \\
Lua 5.4.7 & 1,079 & 600 (55.6\%) & 89 (8.2\%) & 34 & 26 & 26 & 26 & 28 & 28 \\
libuv 1.48.0 & 613 & 335 (54.6\%) & 116 (18.9\%) & 37 & 26 & 27 & 31 & 28 & 31 \\
libpng 1.6.43 & 530 & 237 (44.7\%) & 18 (3.4\%) & 16 & 16 & 16 & 16 & 16 & 16 \\
Binutils libiberty & 622 & 197 (31.7\%) & 165 (26.5\%) & 122 & 115 & 116 & 116 & 115 & 115 \\
BearSSL 0.6 & 605 & 252 (41.7\%) & 53 (8.8\%) & 277 & 263 & 264 & 264 & 264 & 264 \\
OpenSSL 3.5.9 (libssl) & 3,165 & 1,501 (47.4\%) & 392 (12.4\%) & 94 & 73 & 73 & 76 & 73 & 76 \\
OpenSSL 3.5.9 (libcrypto) & 14,674 & 5,440 (37.1\%) & 4,255 (29.0\%) & 980 & 525 & 579 & 630 & 637 & 777 \\
wolfSSL 5.9.4 & 1,322 & 679 (51.4\%) & 161 (12.2\%) & 42 & 33 & 33 & 34 & 33 & 34 \\
Mbed TLS 3.6.4 & 2,174 & 1,103 (50.7\%) & 196 (9.0\%) & 108 & 85 & 87 & 87 & 87 & 87 \\
nginx 1.30.5 & 1,407 & 601 (42.7\%) & 118 (8.4\%) & 120 & 67 & 82 & 85 & 97 & 97 \\
LibreSSL 4.3.3 & 6,716 & 2,718 (40.5\%) & 2,781 (41.4\%) & 480 & 313 & 337 & 374 & 387 & 389 \\
Coreutils 9.5 & 5,819 & 2,798 (48.1\%) & --- & 132 & 75 & 86 & 86 & 86 & 94 \\
curl 8.8.0 & 1,740 & 691 (39.7\%) & 129 (7.4\%) & 213 & 165 & 173 & 174 & 178 & 178 \\
BusyBox 1.36.1 & 2,131 & 647 (30.4\%) & 2,003 (94.0\%) & 493 & 77 & 304 & 311 & 363 & 405 \\
SQLite 3.46.0 & 2,521 & --- & 412 (16.3\%) & 1 & 0 & 0 & 0 & 0 & 0 \\
\midrule
All 16 applications & \multicolumn{3}{c}{} & 3,164 & 1,874 & 2,218 & 2,325 & 2,407 & 2,606 \\
 & \multicolumn{3}{c}{} &  & (59.2\%) & (70.1\%) & (73.5\%) & (76.1\%) & (82.4\%) \\
\bottomrule
\end{tabularx}
\end{table*}
\fi

We start from CompCert-emitted AArch64 assembly at \texttt{-O0} and the
target ABI, with no source ASTs, types, or source-derived signatures. We measure how often reconstruction yields an accepted \ccname{Asm}-to-\ccname{Csyntax} chain. Most of the 3,164 contexts are translation
units. Coreutils uses 132 source-bound contexts from 57 linked images; SQLite
uses one whole-program amalgamation. Each configuration records its search
bounds and receipts.
The 3,173-context oracle in \rq{3} includes nine additional OpenSSL
\texttt{der\_*\_gen.c} data units. They emit no functions or calls, so the
frozen signature campaign has no interface to infer and excludes them.
Table~\ref{tab:no-signature-main}(a) reports exact first-proposal matches
(counts and percentages) out of $N$ held-out CompCert signatures;
(b) reports accepted complete chains out of $N$ configured contexts.
\emph{Only} denotes \tool{} alone, and \emph{Total} denotes the cumulative
engineering portfolio described in \rq{2}. Held-out CompCert signatures are
used only to score proposals after inference, never to generate them.
\tool{} alone accepts 1,874/3,164 contexts (59.2\%). On the shared
14-application scope (36,929 signatures), its first proposal matches 40.9\%
of held-out signatures, compared with 28.6\% for Ghidra. This measures exact
CompCert interface agreement, not the general usefulness of Ghidra's output.

\subsection{\rq{2}: Checked Platform and Decompiler Proposals}

Platform- and decompiler-generated metadata can improve complete-chain coverage by helping \tool{} choose which signature assignments to try. We restrict platform
declarations to checked imports, excluding symbols defined by the image, and
check untrusted Ghidra and IDA/Hex-Rays signatures against independently
decoded AArch64 function entries and calls. Every accepted chain must pass
the same proved validators. Platform proposals add 344 net acceptances,
raising coverage to 70.1\%; the alternative configurations that also use
Ghidra or IDA reach 73.5\% and 76.1\%, respectively. We report the
platform-plus-IDA configuration as our primary assisted result: it accepts
2,407/3,164 contexts and uses no Ghidra input.

The cumulative portfolio builds on the IDA result to reach 2,606 accepted
contexts (82.4\%). Ranking candidates by agreement between Ghidra and IDA
adds 52 contexts to the initial 2,407 acceptances. The remaining gains come
from improved assembly-only reconstruction and further searches for OpenSSL
and Coreutils, including searches guided by dependency profiles.
Public-header proposals include only symbols referenced but not defined by
each input, and some reruns use longer per-stage time bounds; the artifact
records these changes and limits. Because this total combines runs with
different procedures and budgets, we report it as a secondary coverage
result, not a uniform ablation or a per-application maximum over the other
configurations.

An accepted chain need not recover the original interfaces, since the
validators check reconstruction relative to the inferred typed \ccname{Asm}.
In the frozen 15-context zlib evaluation, 104/152 internal-function signatures
and 373/449 call-site signatures match exactly, but only 3/15 accepted
contexts match throughout. Among the 1,198 cumulative acceptances with
comparable archived receipts, 6,809/13,373 internal-function signatures and
44,074/64,802 call-site signatures match exactly, with complete agreement in
112/1,198 contexts. We do not extrapolate from this non-random subset to the
other 1,408 acceptances, which lack comparable receipts;
Appendix~\ref{sec:accepted-signature-fidelity} describes the comparison protocol.
Proposals can reduce the search needed for reconstruction, but time and
candidate limits still constrain coverage. In a libuv rerun with the same
implementation, candidate bound, and timeout, assembly-only inference accepts
26/37 contexts; enabling all hint sources accepts 31/37, reduces checked
candidates from 1,878 to 1,365, and lowers total time from 673.5 to 590.2
seconds. This reduction comes with higher 95th-percentile latency (71.3 to
91.0 seconds); Appendix~\ref{sec:uniform-cost-ablation} reports medians and
counts of functions in accepted translation units. Across the evaluation, candidates
fail validation because of incompatible call shapes, result types, stack
layouts, or complete program interfaces, while some searches exhaust the
candidate bound before finding an accepted assignment. SQLite's 3.9~MB
amalgamation does not finish even its first whole-program candidate within
the time bound. Larger bounds may improve coverage by allowing more search,
but every accepted chain must still satisfy the same validators.

\subsection{\rq{3}: Controlled Exact-Signature Oracle}

We compile the source with CompCert and retain its exact signatures at the
assembly boundary. This oracle isolates reverse-pipeline coverage with ideal
interface information. Table~\ref{tab:application-evaluation} summarizes the
results; the supplement gives the per-application breakdown.

\begin{table}[H]
\centering
\caption{Exact-signature oracle coverage.}
\label{tab:application-evaluation}
\small
\begin{tabular}{@{}llrr@{}}
\toprule
Driver flag & Internal pipeline & Accepted & Eligible \\
\midrule
\texttt{-O0} & unoptimized & 3,173 & 3,173 \\
\texttt{-O1} & optimized (\texttt{-O}) & 3,173 & 3,173 \\
\texttt{-O2} & optimized (same \texttt{-O}) & 3,173 & 3,173 \\
\bottomrule
\end{tabular}
\end{table}

All 3,173 eligible contexts complete the connected
\ccname{Asm}-to-\ccname{Csyntax} chain and recompile with CompCert, covering
approximately 2.16 million physical lines of C. CompCert~3.17 maps literal
\texttt{-O1} and \texttt{-O2} to the same optimized pipeline, so the oracle tests
two CompCert pipelines in total; receipts record each driver invocation. Native Clang~\cite{clang}
\texttt{-O2} tests use Clang's own optimization level. The same 83 configured
inputs are rejected by the CompCert frontend under all flags, before
decompilation.
The smaller \texttt{-O0} separate-pass experiment accepts all 2,696 eligible
files at each of nine reverse boundaries (24,264 local invocations). These
results show that the reverse pipeline handles CompCert's optimized assembly
with supplied interfaces. They do not establish application-scale
signature-free coverage after optimization.

\subsection{\rq{4}: Execution, Generalization, and Output Fidelity}
\label{sec:type-layout-evaluation}

Validator acceptance establishes a semantic relation to the supplied typed
\ccname{Asm}. We also test the C printer, native ABI interoperability, generalization
to unseen inputs, and resemblance to source.
Table~\ref{tab:beyond-acceptance} summarizes execution and generalization;
\emph{Chain/eligible} reports complete connected reconstructions over eligible
contexts (program/driver-flag cells for generated programs).
The large wolfSSL/OpenSSL native studies, the held-out-version study, and the
generated-program study start from retained typed \ccname{Asm} and test the
reverse pipeline, not assembly-only signature inference. The fixed-fixture and
mixed-zlib studies below separately exercise inferred interfaces.

\paragraph{Native execution.}
For wolfSSL and OpenSSL libcrypto at \texttt{-O2}, using retained exact
signatures, we compile original and reconstructed C with both CompCert and
Clang, using the profile in Appendix~\ref{sec:native-clang-profile}. All 42 wolfSSL and 989 OpenSSL contexts
complete the chain. The executables pass 36/36 selected wolfSSL executions
and 836/836 OpenSSL executions. The OpenSSL total comprises 824 upstream-test
executions, four independent-oracle executions, and eight focused
whole-pipeline executions. The upstream tests record 19,900 passing TAP
assertions and no failures. 

\paragraph{Signature-free execution audit.}
On 30 fixed executable fixtures, inferred interfaces yield accepted chains for
27/30, and 23/30 outputs pass the native comparison. Exact interfaces yield
30/30 accepted chains and 29/30 native passes. Three accepted inferred outputs
are not reparsable by CompCert; one further native run needs a Clang adapter.
On the 23 inferred outputs that pass natively, CompCert's interpreter reports
2 expected terminations, 15 internal stuck/undefined-behavior outcomes, 5
unsupported externals, and 1 inline-assembly stop; exact interfaces give the
same split. A current zlib replay accepts 10/15 units; a hybrid retaining the
five original objects passes 3/3 workloads, but is not a full reconstruction.
Appendix~\ref{sec:signature-free-execution-audit} gives the complete audit.

\paragraph{Frozen generalization.}
The producer, profiles, and budgets were frozen before acquiring Lua~5.4.9 and
zlib~1.3.2. All 49 \texttt{-O0} contexts complete the chain and recompile with
CompCert and Clang; original and reconstructed libraries pass 24/24 native
executions. Only 7/49 contexts complete under each optimized flag, so we do
not count the partial libraries as application successes.
Of 288 generated program/flag cells, 267 pass the frontend, 236 complete
reconstruction, and 234 produce matching native records. The remaining two
reconstructed cells are inconclusive: the reference and all comparison
executables time out. No completed comparison produces a mismatch. All 64
bounded programs pass under each of the three flags (192 successful cells).

\begin{table}[H]
\centering
\caption{Native execution and generalization results.}
\label{tab:beyond-acceptance}
\small{}
\setlength{\tabcolsep}{8pt}
\renewcommand{\arraystretch}{1.12}
\begin{tabular}{@{}lrl@{}}
\toprule
\textbf{Cohort} & \textbf{Chain/eligible} & \textbf{Native result} \\
\midrule
wolfSSL \texttt{-O2} & 42/42 & 36/36 executions pass \\
OpenSSL libcrypto \texttt{-O2} & 989/989 & 836/836 executions pass \\
\addlinespace[3pt]
Unseen Lua/zlib \texttt{-O0} & 49/49 & 24/24 executions pass \\
Unseen Lua/zlib optimized & 7/49 & Partial outputs not linked \\
\addlinespace[3pt]
Generated programs & 236/267 & 234 match; 2 inconclusive \\
\addlinespace[3pt]
Fixed ABI fixtures, inferred & 27/30 & 23/30 native comparisons pass \\
Fixed ABI fixtures, exact & 30/30 & 29/30 native comparisons pass \\
Mixed current zlib & 10/15 & 3/3 hybrid workloads pass \\
\bottomrule
\end{tabular}
\end{table}

\paragraph{Output fidelity and type/layout recovery.}
We compare original and reconstructed functions after the same CompCert
frontend elaboration. Across 3,173 accepted contexts and 45,660 matched
function instances, reconstructed C contains 4.11 times as many tokens and
16.96 times as many local declarations as the original. Median
binding-normalized token distance~\cite{wagner1974string} is 0.846; median normalized pq-gram distance~\cite{augsten2010pqgram}
is 0.972 (lower is closer).
The output retains label/goto control flow and
recovers no structured loops or switches. On a fixed 100-function zlib
sample, median raw token distance is 0.845 for \tool{} and 0.691 for default
Ghidra. 
The paired Ascent experiment (Table~\ref{tab:type-layout-main}) uses the same
3,173 inputs and validators. Source types are held out from candidate generation
and ranking. Changes use unrounded counts: relative percentages for counts
and bytes, and percentage points (pp) for agreement and recall.
Definition-sensitive value flow and compatible
layout sharing remove 74,901 casts, 81,320 composite definitions, and 10.35\%
of whole-C bytes. Access-selected physical-field recall rises by 2.41
percentage points, while exact signature-slot agreement falls by 0.16 points.
The analysis often replaces generic pointers with validated physical record
views that differ from the original nominal types.

\begin{table}[H]
\centering
\caption{Paired type/layout refinement.}
\label{tab:type-layout-main}
\small
\setlength{\tabcolsep}{8pt}
\renewcommand{\arraystretch}{1.12}
\begin{tabular}{@{}lrrr@{}}
\toprule
\textbf{Measurement} & \textbf{Previous} & \textbf{Current} & \textbf{Change} \\
\midrule
Function casts & 1,072,844 & 997,943 & $-6.98\%$ \\
Composite definitions & 192,125 & 110,805 & $-42.33\%$ \\
Whole-C bytes & 325,611,708 & 291,905,715 & $-10.35\%$ \\
\midrule
Exact signature-slot agreement & 37.46\% & 37.29\% & $-0.16$ pp \\
Access-selected physical-field recall & 19.59\% & 21.99\% & $+2.41$ pp \\
\bottomrule
\end{tabular}
\end{table}

\section{Related Work}
\label{sec:related-work}

\paragraph{Verified compilation and translation validation.}
CompCert proves semantic preservation for an optimizing compiler by assigning
operational semantics to its intermediate languages and proving simulations
between them~\cite{10.1145/1538788.1538814}. \tool{} uses this infrastructure,
but its theorem does not follow merely by reversing CompCert's theorem: we
prove complementary simulations and compose them with the remaining pass
results. Related CompCert extensions establish separate or open-component
compilation and mixed C/assembly linking
~\cite{stewart2015compositional,koenig2021compcerto,song2020compcertm}; these
techniques are complementary to \tool{}'s connected reverse-pipeline theorem.
Translation validation checks each transformation result instead of
the producer~\cite{pnueli1998translation}. Prior applications include an
optimizing compiler~\cite{necula2000translation}, CompCert instruction
scheduling~\cite{tristan2008validators}, and LLVM optimization validation in
Alive2~\cite{lopes2021alive2}. \tool{} applies the same separation to reverse
transformations: OCaml or Ascent proposes a program and a proved Rocq checker
certifies each accepted result. Other passes are implemented and proved
directly in Rocq, and both forms are composed into one decompiler theorem.

\paragraph{Recompilation-guided recovery.}
Compile-and-compare techniques use a compiler as feedback for search or
evaluation. Evolving Exact Decompilation searches for source programs whose
recompilation matches a target~\cite{schulte2018evolving}. Coda uses
recompilation feedback during iterative recovery~\cite{fu2019coda}, while
LLM4Decompile evaluates reconstructed code by recompilation and
re-execution~\cite{tan2024llm4decompile}. These systems establish useful
syntactic or testing-based checks, but do not provide a composed,
machine-checked semantic theorem for the recovered source. \tool{} uses exact
forward replay where the CompCert pass boundary supports it and proves
complementary simulations where syntactic round trips are inappropriate,
notably across optimized \ccname{RTL}. Uroboros, Ramblr, and Ddisasm instead
target reassemblable or relocatable assembly
~\cite{wang2015uroboros,wang2017ramblr,floresmontoya2020datalog}; they preserve
a lower-level representation rather than certified decompilation.

\paragraph{Recovery of source abstractions.}
Decompilers reconstruct expressions, variables, types, and structured control
flow erased by compilation. Production systems such as Ghidra and RetDec
combine many heuristic analyses to produce usable C-like output
~\cite{ghidra,RetDec}. Cifuentes systematized early
reverse-compilation analyses~\cite{cifuentes1994reverse}; Debin uses learned
scores and approximate inference to recover names and types from stripped
binaries~\cite{he2018debin}; TIE recovers source types with
constraints~\cite{lee2011tie}; and Retypd extends machine-code type
inference with recursive types, polymorphism, and
subtyping~\cite{noonan2016retypd}. TRex instead targets accurate
\emph{behavior-capturing} types rather than an often unrecoverable original
nominal type~\cite{bosamiya2025trex}, a distinction that also motivates
\tool{}'s physical-layout measurements. Phoenix, DREAM, and SAILR recover
structured control flow using structural and compiler-aware
methods~\cite{schwartz2013phoenix,yakdan2015dream,basque2024sailr}.
Recent hybrid systems use learned semantic proposals together with symbolic
reasoning. ReSym fine-tunes separate models for variables and structures, then
uses Prolog to aggregate and cross-check their per-function
predictions~\cite{xie2024resym}. HyRES combines Datalog-based static analysis, an LLM, and
heuristic aggregation to recover structure layouts and semantics from stripped
binaries~\cite{sha2026hyres}. These systems optimize recovery of source-like
names and abstractions; their predictions are not semantic certificates for
the emitted program. \tool{}'s usage of analysis is inspired by these tools, but leverages translation validation to achieve soundness of the result.

\paragraph{Lifting and verified decompilation.}
BAP and angr lift instructions into analysis representations
~\cite{brumley2011bap,shoshitaishvili2016angr}. Formally checked lifters
validate x86-64 instructions or reconstruct sound control-flow
overapproximations~\cite{dasgupta2020lifters,verbeek2022lifting}.
TrABin uses proof-producing ARMv8 and Cortex-M0 transpilers to generate HOL4
certificates for lifting into a binary intermediate language
~\cite{lindner2019trabin}.
Myreen et al. produce logical functions with machine-code
theorems~\cite{myreen2012decompilation}, while BIRD proves a
register-to-variable pass over a generic Coq
representation~\cite{engel2023bird}. FoxDec is the closest effort to sound
recovery at the binary boundary: FoxDec 1.0 recovers restricted,
recompilable C from x86-64 binaries using verified control-flow recovery and
proved symbolic rewrites; its subset requires signatures and headers, excludes
indirect branches, and can require intervention during reference
recovery~\cite{verbeek2020foxdec}. FoxDec 3.0 instead lifts
position-independent x86-64 ELF binaries to symbolized, recompilable
NASM~\cite{verbeek2024lifting}. It recompiles that assembly, independently
lifts both binaries to Ghidra low P-code, and generates
Isabelle/HOL~\cite{nipkow2002isabelle} certificates establishing isomorphic
control-flow graphs and equal corresponding basic-block effects modulo an
address transposition. \tool{} instead consumes textual assembly and
reconstructs recompilable C through a sequence of CompCert intermediate
languages, composing verified passes and proved validators for untrusted
candidate generation into one machine-checked semantic theorem for the complete
reconstruction. Its parser, instruction decoder, and interface reconstruction
are trusted; the theorem
begins with the CompCert \ccname{Asm} program that the frontend constructs.
Connecting that boundary to hardware is complementary work: Sail provides
authoritative Armv8-A ISA semantics, while Islaris verifies machine code
against full-scale ISA semantics~\cite{armstrong2019sail,sammler2022islaris}.

\paragraph{Testing and output fidelity.}
Recompilable output alone does not establish correctness. Empirical studies
find recompilation and semantic failures in C
decompilers~\cite{liu2020decompiler}, while D-Helix compares symbolic models of
lifted and recompiled programs and reports 4,515 incorrect functions across
93,000 Ghidra and angr outputs~\cite{298204}. Conventional tools and \tool{}
therefore do not expose directly comparable success measures: producing
decompiled text is different from returning a complete reconstruction accepted
by a proved validator. \tool{} uses testing to measure coverage, not to justify
correctness. It also separates semantics from resemblance to the unavailable
source: names, types, expressions, and structure can be
lost~\cite{dramko2024taxonomy}. \tool{} certifies one reconstruction without
claiming to recover the original or enumerate every equivalent program.

\section*{AI Usage Statement}

OpenAI's Codex was used as an implementation assistant for our certified Rocq development; under our supervision, we used Codex for code development, proof
repair suggestions, experiment and artifact automation, consistency checks,
and data analysis. The authors wrote the paper manually, but used AI tools for copyediting and critique. The authors take full responsibility for all work.

\section*{Data-Availability Statement}

An anonymized artifact contains \tool{}'s source code, mechanized proofs,
untrusted reconstruction analyses, experiment outputs, paper tables,
and scripts that check the correspondence between them. It provides a pinned
container and separate quick and full validation tracks. The quick track checks
the reported aggregates, focused inference tests, proof hygiene, and one
connected reconstruction; the full track additionally rebuilds and checks the
mechanized proof closure. Third-party application sources that cannot
be redistributed are identified by version with acquisition and preparation
instructions; experiment outputs include hashes where available.
Platform-specific native replays are documented separately from portable proof
and table checks. We intend to submit this artifact for Artifact Evaluation.

\bibliographystyle{ACM-Reference-Format}
\bibliography{references}

\appendix

\section{Detailed Reconstruction and Proofs}\label{sec:appendix}
Sections~\ref{sec:decompilation} and~\ref{sec:formalization} present the central reconstruction problems and the main correctness argument. This
section provides additional technical details on the individual decompilation passes, including their
reconstruction process, formal specifications, and soundness guarantees.
We also present supporting lemmas and proofs underlying the results presented
in the main paper.

\subsection{Decompiler Passes and Their Soundness Proofs}\label{app_sec:formal-proofs}

This section presents the mechanized correctness argument for each semantic
reverse pass. The textual assembly parser and instruction decoder
belong to the trusted front end; the proofs begin with CompCert's unexpanded \ccname{Asm} representation. The proof statements quantify over the unexpanded \ccname{Asm} program and do not require its
function signatures to have been retained from source compilation. The
signature-free front end may therefore propose the typed \ccname{Asm} input without
changing any pass theorem. Its inference remains untrusted, and only
candidates accepted by the validators receive the composed guarantee.

The proof modules for \ccname{Asm}, \ccname{Mach}, and \ccname{Linear} establish the lower reconstruction results. The \ccname{RTL}-to-\ccname{CminorSel} and
\ccname{CminorSel}-to-\ccname{Cminor} reverse-proof modules establish the middle
simulations. The C-front-end validator modules
check the \ccname{Cminor}, \ccname{Csharpminor}, \ccname{Clight}, and \ccname{Csyntax} boundaries, and
\texttt{DecompPipelineproof} composes the accepted programs returned by the
driver. The development reuses CompCert's \ccname{Asmgen}, \ccname{Stacking}, \ccname{Allocation},
\ccname{Cminorgen}, \ccname{Cshmgen}, \ccname{SimplExpr}, and \ccname{SimplLocals} correctness results. Its logical
assumption audit includes CompCert's selected target configuration and
external-semantics interfaces, standard extensionality and proof-irrelevance
principles used by CompCert, and classical logic used by the reverse proofs.
There are no admitted proof obligations or correctness axioms for the
candidate generators. Parsing, printing, signature inference, assembling,
linking, and native execution remain outside the mechanized theorem.

One computational constant also appears in the audit and should not be
mistaken for a generator-correctness assumption.
\texttt{DecompLTL\_Validation.decompile\_ltl\_function} is declared as a
Rocq \texttt{Parameter} of computational type because the development does
not execute the untrusted OCaml generator during proof checking. Extraction
replaces this constant with the implementation in \texttt{DecompLTL}. No
proposition about the result of that function is assumed: the theorem is
conditional only on the verified allocation validator accepting the proposed
\ccname{RTL} function. Consequently, an arbitrary or faulty replacement can
reduce coverage or fail validation, but cannot establish a false accepted
result.

\subsection{\ccname{Asm} to \ccname{Mach}}
The \ccname{Asm}-to-\ccname{Mach} decompiler is implemented in OCaml as a syntax-directed inverse of CompCert's $\mathsf{Asmgen}$ pass. It recognizes the target instruction sequences generated for \ccname{Mach} operations, calls, branches, returns, builtins, and stack-frame operations, and reconstructs the corresponding \ccname{Mach} instructions while preserving labels and control flow. Because some assembly patterns admit several possible reconstructions, the decompiler may generate multiple candidates. Each candidate is checked by rerunning $\mathsf{Asmgen}$, and only a candidate that reproduces the complete supplied unexpanded \ccname{Asm} program is accepted.

The textual-assembly parser and the recovery of symbols and function
signatures precede this proved boundary. Signatures can either be retained by
CompCert or inferred from assembly and the target ABI. In both cases, the
semantic theorem is quantified over the resulting typed \ccname{Asm} program. The
validator certifies the \ccname{Mach} reconstruction relative to this typed
\ccname{Asm} candidate; fidelity of inferred signatures to external assembly
remains a trusted front-end obligation.

The \ccname{Asm}-to-\ccname{Mach} validator reruns CompCert's \ccname{Asmgen} pass on the
reconstructed \ccname{Mach} program. It accepts only if the result is exactly the
supplied unexpanded \ccname{Asm} program:
\[
  \mathsf{validate}(P_A,P_M)=\mathsf{true}
  \Longrightarrow
  \mathsf{Asmgen}(P_M)=\mathsf{OK}(P_A).
\]

We first define decidable equality for \ccname{Asm} instructions and operands, including registers, addressing modes, labels, builtins, frame operations, and branch targets. We extend this equality to functions and complete programs, then apply \texttt{Asmgenproof.transf\_program\_correct}. The proof reuses CompCert's target-specific return-address relation.

\begin{theorem}[\ccname{Asm}-to-\ccname{Mach} semantic preservation]
Let $P_A$ be an unexpanded \ccname{Asm} program and $P_M$ an accepted \ccname{Mach}
reconstruction. Then
\[
\begin{aligned}
  &\mathsf{validate}(P_A,P_M)=\mathsf{true}
  \land P_M\Downarrow B_M
  \Longrightarrow
  \exists B_A.\;
  P_A\Downarrow B_A
  \land B_M\preceq B_A.
\end{aligned}
\]
\end{theorem}
\paragraph{Proof sketch.}
The validator reruns CompCert's $\mathsf{Asmgen}$ transformation on the
reconstructed \ccname{Mach} program and accepts only when
$\mathsf{Asmgen}(P_M)=\mathsf{OK}(P_A).$
Validator soundness follows by unfolding the executable check and using
decidable equality of complete \ccname{Asm} programs. We then apply CompCert's existing
$\mathsf{Asmgen}$ correctness theorem, which gives a forward simulation from
the semantics of $P_M$ to the semantics of $P_A$. CompCert's generic simulation
theorem converts this simulation into the behavior-improvement result
$B_M\preceq B_A$. Finally, when $B_M$ is safe, behavior improvement cannot
replace it with an undefined behavior; therefore $B_A=B_M$.

\begin{corollary}[Preservation of defined \ccname{Mach} behaviors]
\[
\begin{aligned}
  &\mathsf{validate}(P_A,P_M)=\mathsf{true}
  \land P_M\Downarrow B
  \land \mathsf{Safe}(B)
  \Longrightarrow P_A\Downarrow B.
\end{aligned}
\]
\end{corollary}
\paragraph{Proof sketch.}
The corollary follows directly from the theorem. Since $\mathsf{Safe}(B)$ excludes undefined behavior,
$B_M\preceq B_A$ implies $B_A=B_M=B$. Hence the supplied \ccname{Asm} program
exhibits the same behavior $B$.

\subsection{\ccname{LTL} to \ccname{RTL}}
CompCert's register allocation transforms \ccname{RTL} into \ccname{LTL} by replacing pseudo-registers with physical registers and stack slots, introducing moves, spills, and reloads when necessary. Our \ccname{LTL}-to-\ccname{RTL} decompiler reverses this transformation by reconstructing pseudo-registers and eliminating instructions introduced by register allocation. We follow a translation-validation approach: an untrusted OCaml decompiler generates an \ccname{RTL} candidate, whose correctness is checked using CompCert's register-allocation validator.

\begin{figure}[tbp]
\centering
\captionsetup{skip=5pt}

\begin{minipage}[t]{0.27\linewidth}
\codeheading{C program}
\begin{lstlisting}[style=code,basicstyle=\ttfamily\footnotesize]
int add(int x, int y) {
  return x + y;
}
\end{lstlisting}
\end{minipage}%
\codegap%
\begin{minipage}[t]{0.25\linewidth}
\codeheading{\ccname{LTL}}
\begin{lstlisting}[style=code,basicstyle=\ttfamily\footnotesize]
3:  X19      := X0
    stack[0] := X1
    goto 2

2:  X20 := stack[0]
    X0  := X19 + X20
    goto 1

1:  return
\end{lstlisting}
\end{minipage}%
\codegap%
\begin{minipage}[t]{0.25\linewidth}
\codeheading{\ccname{RTL}}
\begin{lstlisting}[style=code,basicstyle=\ttfamily\footnotesize]
parameters: r1, r2

2:  r3 := r1 + r2
    goto 1

1:  return r3
\end{lstlisting}
\end{minipage}

\caption{Compilation of \textsf{add} to \ccname{RTL} and \ccname{LTL}.}
\Description{Side-by-side control-flow graphs showing pseudo-register RTL and
physical-register plus stack-slot LTL for an integer addition.}
\label{fig:rtl-ltl-example}
\end{figure}

Figure~\ref{fig:rtl-ltl-example} illustrates the relationship between \ccname{RTL} and \ccname{LTL} using a simple function, \texttt{add}, which returns the sum of two integers. For simplicity, we illustrate register spilling and reloading using this small example, although register allocation for such a function would typically not require these operations. In the \ccname{LTL} representation, \texttt{X0} and \texttt{X1} initially hold the arguments \texttt{x} and \texttt{y}, respectively, while \texttt{X0} is also used to return the result. In block 3, the value of \texttt{x} is moved to register \texttt{X19}, and the value of \texttt{y} is spilled into an abstract stack slot. In block 2, \texttt{y} is reloaded into register \texttt{X20}, and the addition is performed, storing the result in \texttt{X0}.

The corresponding \ccname{RTL} representation is simpler. It uses pseudo-registers \texttt{r1} and \texttt{r2} for the arguments and \texttt{r3} for the result, without explicitly managing physical registers or stack slots. Reconstructing \ccname{RTL} from \ccname{LTL} therefore requires identifying and eliminating instructions introduced by register allocation, such as moves, spills, and reloads, while recovering suitable pseudo-registers and instructions. For example, the sequence of instructions in \ccname{LTL} blocks 3 and 2 can be reconstructed as a single \ccname{RTL} addition instruction.

\paragraph{Overview of decompilation.}
The decompiler maintains an environment
$E : \mathit{location} \rightharpoonup \mathit{register}$,
where a location represents a physical register or stack slot,
and a register represents a reconstructed \ccname{RTL} pseudo-register.
Multiple locations may map to the same pseudo-register because
register allocation can move or copy a value between locations.
The function signature and target calling convention determine
the initial locations of function parameters.

Consider the example in Figure~\ref{fig:rtl-ltl-example}.
Initially, the environment maps $\texttt{X0}$ to $\texttt{r1}$
and $\texttt{X1}$ to $\texttt{r2}$.
The instruction $\texttt{X19 := X0}$ updates the environment
with $E(\texttt{X19}) = \texttt{r1}$.
Similarly, spilling $\texttt{X1}$ into $\texttt{stack[0]}$
adds $E(\texttt{stack[0]}) = \texttt{r2}$.
When $\texttt{X20 := stack[0]}$ reloads the value,
the environment is updated with
$E(\texttt{X20}) = \texttt{r2}$. The reconstruction algorithm first processes the moves
preceding an \ccname{LTL} computation and updates the environment.
It then reconstructs the corresponding \ccname{RTL} instruction,
introducing a fresh pseudo-register for the computed result. Finally, it processes any trailing moves and identifies the successor node. Thus, by traversing the \ccname{LTL} control-flow graph and tracking how values move between physical locations, the decompiler
recovers the pseudo-registers and reconstructs the \ccname{RTL}
instructions while eliminating moves introduced by
register allocation.

\paragraph{Translation validation.}
The reconstruction algorithm is implemented in OCaml and is not
part of the trusted proof. Each generated \ccname{RTL} candidate $P_R$ is
checked against the original \ccname{LTL} program $P_L$ using two existing
CompCert components. First, \texttt{RTLtyping.type\_function}
checks that the reconstructed \ccname{RTL} function is well typed.
Second, \texttt{Allocation.check\_function} validates that the
\ccname{LTL} function corresponds to a valid register allocation of the
reconstructed \ccname{RTL} function. This includes checking the
correspondence of operations, register locations, spills, reloads,
control-flow edges, calls, and returns. For the example in Figure~\ref{fig:rtl-ltl-example}, validation
establishes that pseudo-register \texttt{r1} corresponds to
\texttt{X0} and subsequently \texttt{X19}, while \texttt{r2}
corresponds to \texttt{X1}, \texttt{stack[0]}, and \texttt{X20}.
It also checks that the \ccname{LTL} addition implements the reconstructed
\ccname{RTL} addition and that the result is returned through \texttt{X0}.
Candidates that fail either check are rejected.
The correctness of \ccname{LTL}-to-\ccname{RTL} decompilation follows the
translation-validation approach described in
Section~\ref{sec:decompilation}. Successful \ccname{RTL} type checking
and register-allocation validation establish CompCert's
allocation specification for the reconstructed \ccname{RTL} program
$P_R$ and the supplied \ccname{LTL} program $P_L$.
We therefore reuse CompCert's existing correctness theorem
for validated register allocation, which establishes a
forward simulation from $P_R$ to $P_L$.
Consequently, every defined observable behavior of the reconstructed $P_R$
is also a behavior of the supplied $P_L$.

The central validator lemma is
\[
 \mathsf{validate\_function}(f_T,f_R)=\mathsf{OK}
  \Longrightarrow
  \mathsf{transf\_function\_spec}(f_R,f_T).
\]
Its proof unfolds the validator and applies two existing results. CompCert's
\ccname{RTL} type-checking theorem proves the inferred register typing. Its allocation
checker soundness theorem converts successful checking into the semantic
register-allocation relation.

This function relation is lifted to entire programs. Internal functions
satisfy the checked allocation relation; external functions, globals, public
symbols, and the main symbol are preserved. The final theorem is obtained by
instantiating \texttt{Allocproof.validated\_program\_correct}.

\begin{theorem}[\ccname{LTL}-to-\ccname{RTL} semantic preservation]
Let $P_T$ be an \ccname{LTL} program and let \\
$\mathcal{D}_{\mathsf{LTL\rightarrow RTL}}(P_T)
=\mathsf{OK}(P_R)$. Then
$\begin{aligned}
  &P_R\Downarrow B_R
  \Longrightarrow
  \exists B_T.\;
  P_T\Downarrow B_T
  \land B_R\preceq B_T.
\end{aligned}
$
\end{theorem}

\begin{corollary}[Preservation of defined \ccname{RTL} behaviors]
$$
\begin{aligned}
  &\mathcal{D}_{\mathsf{LTL\rightarrow RTL}}(P_T)
       =\mathsf{OK}(P_R)
  \land P_R\Downarrow B
  \land \mathsf{Safe}(B)
  \Longrightarrow P_T\Downarrow B.
\end{aligned}
$$
\end{corollary}

The state relation used by CompCert's register allocation validator allows an \ccname{RTL} register to hold a value less defined than that of its corresponding \ccname{LTL} location. Consequently, the correctness theorem establishes a forward simulation from the reconstructed \ccname{RTL} program to the original \ccname{LTL} program, ensuring that the reconstruction introduces no new defined observable behaviors. However, this guarantee is directional and does not establish the converse simulation for full behavioral equivalence.

\subsection{\ccname{RTL} to \ccname{CminorSel}}
\ccname{RTL}-to-\ccname{CminorSel} reconstruction is implemented and verified directly in Rocq. The proof is organized at the expression, instruction, function, and program levels. First, a register-environment relation states that every \ccname{RTL} register used by the function has the same value as the corresponding \ccname{CminorSel} local variable. The relation is restricted to the finite set of registers occurring in the function. \ccname{RTL} register maps assign \texttt{Vundef} to uninitialized registers, whereas the \ccname{CminorSel} environment contains only declared parameters and
locals. The relation permits corresponding values to be \texttt{Vundef}; it does not assume that every register contains a defined value. We prove the following properties: 
\begin{itemize}
  \item a reconstructed register expression evaluates to the corresponding
        \ccname{RTL} register value;
  \item reconstructed operation operands preserve their order and values;
  \item reconstructed addressing expressions compute the same memory address;
  \item call targets and argument lists denote the same functions and values.
\end{itemize}

The proof considers each \ccname{RTL} instruction. Assignments preserve the correspondence between registers and local variables, while loads and stores use addressing lemmas to establish the same memory effects. Conditional branches select the same successor. The remaining cases handle calls, tail calls, builtins, jump tables, and returns. The state relation connects an \ccname{RTL} state at program counter $pc$ with a \ccname{CminorSel} state executing \texttt{goto pc}. Label lookup identifies the corresponding reconstructed statement. The relation also connects \ccname{RTL} call stacks with \ccname{CminorSel} continuations and ensures that their register and local environments agree. The state relation also includes an \ccname{RTL} value-analysis invariant that justifies simplifying reconstructed expressions and conditions. We prove that this invariant holds initially and is preserved by every \ccname{RTL} step, without requiring all registers to contain defined values. Initial states are related through preservation of global symbols, and related final states return the same integer value.

The step theorem shows that each \ccname{RTL} transition is matched by one or more \ccname{CminorSel} transitions with the same observable trace. Multiple target steps are needed because one \ccname{RTL} instruction can reconstruct to a sequence of
\ccname{CminorSel} assignments, statements, labels, and gotos. The complete simulation is therefore proved using CompCert's
\texttt{forward\_simulation\_plus} construction.

\begin{theorem}[\ccname{RTL}-to-\ccname{CminorSel} semantic preservation]
Let $P_R$ be an \ccname{RTL} program and let
$
  \mathcal{D}_{\mathsf{RTL\rightarrow CminorSel}}(P_R)
  =\mathsf{OK}(P_S).
$
Then
$
  P_R\Downarrow B_R
  \Longrightarrow
  \exists B_S.\;
  P_S\Downarrow B_S
  \land B_R\preceq B_S.
$
\end{theorem}
\paragraph{Proof sketch.}
Successful reconstruction establishes correspondence between the global
environments, functions, \ccname{RTL} nodes, and reconstructed \ccname{CminorSel} statements.
The expression lemmas show that corresponding operands, operations, addresses,
conditions, and call arguments evaluate to the same values. Using these
lemmas, the instruction proof shows that every \ccname{RTL} transition is reproduced by
one or more \ccname{CminorSel} transitions with the same trace, memory effects, and
updated environment. Initial-state and final-state correspondence are combined
with this step result using \texttt{forward\_simulation\_plus}. CompCert's
generic behavior theorem then produces a \ccname{CminorSel} behavior $B_S$ satisfying
$B_R\preceq B_S$.

\begin{corollary}[Preservation of defined \ccname{RTL} behaviors]
$$
\begin{aligned}
  &\mathcal{D}_{\mathsf{RTL\rightarrow CminorSel}}(P_R)
       =\mathsf{OK}(P_S)
  \land P_R\Downarrow B
  \land \mathsf{Safe}(B)
  \rightarrow P_S\Downarrow B.
\end{aligned}
$$
\end{corollary}

\paragraph{Proof sketch.}
For the corollary, $\mathsf{Safe}(B)$ excludes undefined \ccname{RTL} behavior.
Therefore $B\preceq B_S$ implies $B_S=B$, and the reconstructed \ccname{CminorSel}
program exhibits the same behavior $B$.

\paragraph{Complementary simulation for composition.}
The preceding simulation follows an \ccname{RTL} execution through the reconstructed
\ccname{CminorSel} program. To obtain a continuous semantic path from reconstructed C
to the supplied assembly, we also prove the complementary simulation from the
reconstructed \ccname{CminorSel} program to the supplied \ccname{RTL} program.
The main difficulty is that one \ccname{RTL} node is reconstructed as a \ccname{CminorSel}
statement containing assignments, sequences, labels, blocks, and gotos. Some
\ccname{CminorSel} transitions therefore perform only administrative work and have no
corresponding \ccname{RTL} transition. The complementary proof uses a phase-indexed
state relation. A related state contains a \ccname{CminorSel} state, its corresponding
\ccname{RTL} state, and a natural-number index representing the remaining
administrative phases. A \ccname{CminorSel} transition that performs the action reconstructed from an \ccname{RTL}
instruction is matched by that \ccname{RTL} instruction with the same observable
trace. A transition that only traverses a sequence, label, block, or goto may
leave the \ccname{RTL} state unchanged, but it must strictly decrease the phase index.
Because the index is a natural number, these silent transitions cannot
continue indefinitely.

Inverse-evaluation lemmas show that evaluating a reconstructed \ccname{CminorSel}
expression recovers the corresponding \ccname{RTL} register values, operation
operands, memory addresses, conditions, call targets, and jump-table
destinations. Successful reconstruction also supplies the required
correspondence between entry nodes, successor nodes, symbols, functions, and
calling conventions. As in the preceding direction, the relation includes
the proved \ccname{RTL} value-analysis invariant needed to justify reconstructed
normalizations.

\begin{theorem}[Complementary \ccname{CminorSel}-to-\ccname{RTL} simulation]
Let $P_R$ be an \ccname{RTL} program and let
$\mathcal{D}_{\mathsf{RTL\rightarrow CminorSel}}(P_R)
  =\mathsf{OK}(P_S).$
Then
$\mathsf{forward\_simulation}
    \bigl(\mathsf{Sem}_{\mathsf{CminorSel}}(P_S),
          \mathsf{Sem}_{\mathsf{RTL}}(P_R)\bigr).
$

Consequently,
$P_S\Downarrow B_S
  \Longrightarrow
  \exists B_R.\;
  P_R\Downarrow B_R
  \land B_S\preceq B_R.
$
\end{theorem}

\paragraph{Proof sketch.}
The proof relates each \ccname{CminorSel} execution phase to the \ccname{RTL} instruction from
which it was reconstructed. For every \ccname{CminorSel} transition, the \ccname{RTL} program
either executes the corresponding instruction with the same trace or remains
unchanged while the phase index strictly decreases. Initial states are related
using preservation of memory, symbols, the main function, and its signature;
related final states return the same integer value. We use the well-founded
order on natural numbers and determinacy of \ccname{CminorSel} semantics to assemble
these results with CompCert's
\texttt{forward\_simulation\_determ} construction.

This theorem requires no additional semantic assumption. Together with the
\ccname{RTL}-to-\ccname{CminorSel} theorem above, it establishes both simulation directions for
every successfully reconstructed pair at this boundary.

\subsection{\ccname{CminorSel} to \ccname{Cminor}}
\ccname{CminorSel} expressions can contain target-specific operations and value-level
selections that are absent from \ccname{Cminor}. Reconstruction may therefore produce a
sequence of setup statements followed by a simpler \ccname{Cminor} expression.

The expression proof is organized around a relation
$\mathsf{transl\_expr}(e_S)
     =\mathsf{OK}(s_C,e_C),
$
where $s_C$ computes any required temporary values and $e_C$ denotes the final
result. The central expression lemma shows that evaluating $e_S$ in a
\ccname{CminorSel} environment is reproduced by executing $s_C$ and then evaluating
$e_C$ in the resulting \ccname{Cminor} environment. Supporting lemmas cover constants,
variables, operations, addressing modes, loads, conditions, argument lists,
and call targets. Freshness lemmas ensure that setup statements modify only
newly introduced temporaries.

The proof then proceeds by induction over \ccname{CminorSel} statements. Sequencing is
handled by composing the corresponding \ccname{Cminor} executions. Conditions and
selections use the expression theorem before entering the reconstructed branch.
Loops, blocks, and exits preserve their continuation structure. Further cases
cover calls, tail calls, builtins, switches, labels, gotos, and returns.

Three side conditions are required by the current proof:
\begin{enumerate}
  \item every reached unsigned-long comparison produces a defined Boolean
        result;
  \item every reached selection has a defined condition; and
  \item every reached function contains unique labels.
\end{enumerate}
The first two conditions account for a semantic difference: \ccname{CminorSel} may
propagate \texttt{Vundef} through a value-level operation, while reconstructed
\ccname{Cminor} control flow can become stuck when asked to branch on an undefined
condition. The third condition is used because expanding a nested selection
can duplicate a labeled branch; uniqueness ensures that a reconstructed
\texttt{goto} still has an unambiguous corresponding target.

The state relation combines expression-environment agreement with
corresponding call stacks, functions, statements, and memories. A strengthened
invariant carries label uniqueness across execution. Every \ccname{CminorSel} step is
matched by one or more \ccname{Cminor} steps, after which the resulting states again
satisfy the invariant.

\begin{theorem}[\ccname{CminorSel}-to-\ccname{Cminor} semantic preservation]
Let $P_S$ be a \ccname{CminorSel} program and let
$\mathcal{D}_{\mathsf{CminorSel\rightarrow Cminor}}(P_S)
=\mathsf{OK}(P_C)$. If reached unsigned-long comparisons and selections are
defined and reached functions have unique labels, then
$
  P_S\Downarrow B_S
  \Longrightarrow
  \exists B_C.\;
  P_C\Downarrow B_C
  \land B_S\preceq B_C.
$
\end{theorem}

\begin{corollary}[Preservation of defined \ccname{CminorSel} behaviors]
Under the same side conditions,
$
  \mathcal{D}_{\mathsf{CminorSel\rightarrow Cminor}}(P_S)
       =\mathsf{OK}(P_C)
  \land P_S\Downarrow B
  \land \mathsf{Safe}(B)
  \Longrightarrow P_C\Downarrow B.
$
\end{corollary}

These three conditions belong to the original pass-local simulation from
\ccname{CminorSel} to \ccname{Cminor}. They are not premises of the complete decompiler theorem,
which uses the complementary simulation below at this boundary.

\paragraph{Complementary simulation for composition.}
For end-to-end composition we also require a simulation from the reconstructed
\ccname{Cminor} program to its \ccname{CminorSel} input.  This direction is proved for the
\ccname{CminorSel} programs generated by the preceding \ccname{RTL}-to-\ccname{CminorSel} pass.  This
restriction is significant: arbitrary \ccname{CminorSel} permits effectful operations
inside expressions, whereas \ccname{RTL}-generated \ccname{CminorSel} uses an effect-free
expression subset whose effects occur at statement boundaries.  The
successful \ccname{RTL} reconstruction provides a provenance witness showing the
exact \ccname{RTL} instruction and execution phase represented by every reached
\ccname{CminorSel} state.

Reconstruction of one \ccname{CminorSel} expression can introduce several silent
\ccname{Cminor} assignments before its observable action and additional silent
bookkeeping afterward.  The state relation therefore has three phases:
\emph{before}, while setup precedes the matching action; \emph{after}, while
only trailing administration remains; and \emph{stuck}, for a finite silent
suffix whose reconstructed computation is undefined.  Each phase carries a
natural-number rank.  At exact instruction boundaries the relation combines
the ordinary \ccname{CminorSel}-to-\ccname{Cminor} match relation with generated-syntax and \ccname{RTL}
provenance invariants.  Constructor-specific lemmas cover operations, loads,
stores, calls, builtins, conditions, jump tables, gotos, and returns.

The syntax-specific step proof is first established for safe \ccname{Cminor} states,
where progress permits inverse evaluation of every generated setup
expression.  This local use of safety does not become a theorem premise.  If
a related \ccname{Cminor} state is unsafe, negating CompCert's definition of safety
exposes a finite silent path to a non-final state with no transition; the
\emph{stuck} phase simulates that path by stuttering with a decreasing rank.
Observable external and builtin transitions are replayed in \ccname{CminorSel} with the
same trace, using receptiveness where the external result is chosen by the
environment.  Recursive progress lemmas handle structured builtin arguments.
For an unsigned-long comparison or value-level selection, the successful
\ccname{Cminor} transition being simulated supplies the defined Boolean result needed
by the corresponding \ccname{CminorSel} operation. If the condition is undefined,
\ccname{Cminor} has no transition, so there is no source step to match. Setup statements
generated from \ccname{RTL} operands contain only variable reads and are covered by
separate total lemmas. Consequently, these partial operations require no
global definedness hypothesis in the complementary direction.

\begin{theorem}[Complementary \ccname{Cminor}-to-\ccname{CminorSel} simulation]
Let $P_R$, $P_S$, and $P_C$ be \ccname{RTL}, \ccname{CminorSel}, and \ccname{Cminor} programs such that
\[
\begin{aligned}
  \mathcal{D}_{\mathsf{RTL\rightarrow CminorSel}}(P_R)
    &=\mathsf{OK}(P_S),\\
  \mathcal{D}_{\mathsf{CminorSel\rightarrow Cminor}}(P_S)
    &=\mathsf{OK}(P_C).
\end{aligned}
\]
Then
$
  \mathsf{forward\_simulation}
    \bigl(\mathsf{Sem}_{\mathsf{Cminor}}(P_C),
          \mathsf{Sem}_{\mathsf{CminorSel}}(P_S)\bigr).
$
Consequently,
\[
  P_C\Downarrow B_C
  \Longrightarrow
  \exists B_S.\;
  P_S\Downarrow B_S
  \land B_C\preceq B_S.
\]
\end{theorem}
\paragraph{Proof sketch.}
Successful \ccname{RTL}-to-\ccname{CminorSel} and \ccname{CminorSel}-to-\ccname{Cminor} reconstruction establishes
the generated-syntax, label-uniqueness, register-environment, and \ccname{RTL}
provenance invariants at each instruction boundary. The proof relates \ccname{Cminor}
and \ccname{CminorSel} states using the \emph{before}, \emph{after}, and \emph{stuck}
phases, ordered by a well-founded natural-number rank.

For a safe \ccname{Cminor} state, inversion of the \ccname{Cminor} transition and the
constructor-specific evaluation lemmas identify the corresponding \ccname{CminorSel}
action. Silent setup and cleanup steps are handled by decreasing the phase
rank, while observable calls and builtins are replayed with the same trace.
For an unsafe \ccname{Cminor} state, CompCert's definition of safety yields a finite
silent execution to a non-final state with no successor. The \emph{stuck}
phase represents this execution while decreasing the rank.
Initial states are related using preservation of global environments and
function entry, and related final states return the same value. These results
are combined with the well-founded phase order to construct the forward
simulation. Because safe and unsafe states are both handled inside the proof,
the final theorem requires no global safety or definedness assumption.

\subsection{\ccname{Cminor} to \ccname{Csharpminor}}
The \ccname{Cminor}-to-\ccname{Csharpminor} generator represents each nonempty stack frame as one addressable object and reconstructs interior byte displacements from its
base. Let $\mathsf{Frame}(P_H)$ require every function in $P_H$ to contain at most one addressable object and require the size of that object to be nonnegative. Let $N_s$ denote the normalization that folds constant \ccname{Cminor} stack-address displacements. Validator soundness establishes
\[
\begin{aligned}
  &\mathsf{validate}(P_C,P_H)=\mathsf{true}
  \rightarrow
  \mathsf{Frame}(P_H)
  \land
  \exists P_G.\;
    \mathsf{Cminorgen}(P_H)=\mathsf{OK}(P_G)\\
  &\hspace{5.5em}\land
    \bigl(P_G=P_C
       \lor N_s(P_G)=N_s(P_C)\bigr).
\end{aligned}
\]
The comparison includes complete \ccname{Cminor} programs: function bodies, external
declarations, global variables, public symbols, their order, and the program
entry point. The frame-shape check is required even when the unnormalized programs match.
Without it, a candidate could divide one \ccname{Cminor} stack block into several \ccname{Csharpminor} objects. \ccname{Cminorgen} could assign those objects contiguous offsets and reproduce the same \ccname{Cminor} syntax, although the reconstructed program would
have different object boundaries and could interpret interior accesses differently. In the normalized case, \texttt{CminorStackAddr} provides a simulation from
$P_G$ to $N_s(P_G)$ and a simulation from $N_s(P_C)$ back to $P_C$. Composing these simulations with
\texttt{Cminorgenproof.transl\_program\_correct} gives the same semantic guarantee as exact round-trip validation.

\begin{theorem}[\ccname{Cminor}-to-\ccname{Csharpminor} semantic preservation]
Let $P_C$ be a \ccname{Cminor} program and $P_H$ an accepted \ccname{Csharpminor}
reconstruction. Then
\[
\begin{aligned}
  &\mathsf{validate}(P_C,P_H)=\mathsf{true}
  \land P_H\Downarrow B_H
  \rightarrow
  \exists B_C.\;
  P_C\Downarrow B_C
  \land B_H\preceq B_C.
\end{aligned}
\]
\end{theorem}

\paragraph{Proof sketch.}
Unfolding the validator yields the frame-shape property, a generated \ccname{Cminor}
program $P_G$, and either exact equality $P_G=P_C$ or normalized equality
$N_s(P_G)=N_s(P_C)$. In the exact case, CompCert's \ccname{Cminorgen} correctness
theorem directly gives a forward simulation from $P_H$ to $P_C$.
In the normalized case, we first apply \ccname{Cminorgen} correctness to obtain a
simulation from $P_H$ to $P_G$. We then compose it with the proved simulation
from $P_G$ to $N_s(P_G)$. Normalized equality identifies this program with
$N_s(P_C)$, and the reverse normalization theorem supplies the final simulation
from $N_s(P_C)$ to $P_C$. CompCert's generic behavior theorem therefore
produces a \ccname{Cminor} behavior $B_C$ satisfying $B_H\preceq B_C$.

\begin{corollary}[Preservation of defined \ccname{Csharpminor} behaviors]
\[
\begin{aligned}
  &\mathsf{validate}(P_C,P_H)=\mathsf{true}
  \land P_H\Downarrow B
  \land \mathsf{Safe}(B)
  \rightarrow P_C\Downarrow B.
\end{aligned}
\]
\end{corollary}
$\mathsf{Safe}(B)$ excludes undefined \ccname{Csharpminor} behavior.
Therefore $B\preceq B_C$ implies $B_C=B$, and the supplied \ccname{Cminor} program has
the same behavior $B$.

\subsection{\ccname{Csharpminor} to \ccname{Clight}}
The OCaml extractor records direct type facts from function signatures,
operations, memory accesses, calls, builtins, and assignments. Ascent then
propagates these facts across copy-related variables. The central relations
are
\[
\mathsf{evidence}(x,\tau),\qquad
\mathsf{equal}(x,y),\qquad
\mathsf{sameType}(x,y),\qquad
\mathsf{inferred}(x,\tau).
\]
Copy edges generate symmetric same-type edges; their transitive closure then
propagates inferred types:
\[
\begin{gathered}
\overset{\text{\textsc{T-Copy}}}
  {\frac{\mathsf{equal}(x,y)}{\mathsf{sameType}(x,y)}}
\qquad
\overset{\text{\textsc{T-Sym}}}
  {\frac{\mathsf{equal}(x,y)}{\mathsf{sameType}(y,x)}}
\qquad
\overset{\text{\textsc{T-Trans}}}
  {\frac{\mathsf{sameType}(x,y)\quad\mathsf{sameType}(y,z)}
         {\mathsf{sameType}(x,z)}}
\\[2ex]
\overset{\text{\textsc{T-Evidence}}}
  {\frac{\mathsf{evidence}(x,\tau)}{\mathsf{inferred}(x,\tau)}}
\qquad
\overset{\text{\textsc{T-Propagate}}}
  {\frac{\mathsf{inferred}(x,\tau)\quad\mathsf{sameType}(x,y)}
         {\mathsf{inferred}(y,\tau)}}.
\end{gathered}
\]
For example, in \texttt{t = x; result = Odivu(t,y)}, unsigned division suggests
unsigned-integer types for \texttt{t}, \texttt{y}, and \texttt{result};
the copy relation assigns the same candidate type to \texttt{x}.
The analysis retains conflicting type candidates as alternatives.

Addressable variables retain their size in bytes but not their element type.
A 40-byte object accessed with 32-bit loads can therefore be proposed as
\texttt{int buffer[10]}. When the access patterns do not suggest an element
type, the decompiler proposes a byte array. Scaled addresses such as
$\mathsf{\&buffer}+\mathit{index}\times 4$ can become
$\mathsf{buffer[index]}$ when the inferred element size is four bytes. The
scale must be removed during reconstruction because \ccname{Clight} reintroduces it
when compiling the array access.

\ccname{Csharpminor} conditionals are already structured, while combinations of
\texttt{Sblock}, \texttt{Sloop}, and numbered \texttt{Sexit} encode loops,
\texttt{break}, and \texttt{continue}. The generator recognizes these forms
and blocked switch structures, but interprets a numbered exit as a structured
statement only when the surrounding nesting justifies it. Control flow that
does not match these patterns remains explicit.

These facts guide candidate generation but are not trusted by the correctness
proof. An incorrect type, layout, or structure inference may cause rejection,
but it cannot establish the theorem for an incorrect candidate.

Let $P_C$ denote the supplied \ccname{Csharpminor} program and let $P_H$ denote its
reconstructed \ccname{Clight} program. The validator first checks that every variadic
prototype in $P_H$ contains exactly the fixed parameter prefix recorded by its
calling convention. It then recompiles $P_H$ using CompCert's
\ccname{Cshmgen} pass and compares the complete generated \ccname{Csharpminor} program
with $P_C$ and a finite set of permitted normalizations of $P_C$.

We write $\mathcal{N}(P_C)$ for this set. It contains $P_C$ itself and the
programs obtained by composing the following proved transformations:
discarded-result cleanup, guarded narrow-result restoration, canonical-result
restoration, narrow-assignment restoration, and parameter-entry restoration.
More precisely, let $U$, $N$, $C$, $A$, and $P$ denote these transformations,
respectively. Then
\[
\begin{aligned}
\mathcal{N}(P_C)=
\bigl\{\,
  &P_C,\;
  U(P_C),\;
  P(N(U(P_C))),\;
  P(N(C(U(P_C)))),\\
  &P(A(N(U(P_C)))),\;
  P(A(N(C(U(P_C)))))
\,\bigr\}.
\end{aligned}
\]

Validator soundness establishes
\[
\begin{aligned}
  &\mathsf{validate}(P_C,P_H)=\mathsf{true}
  \rightarrow
    \mathsf{Arity}(P_H)
    \land
    \exists P_G.\;
      \mathsf{Cshmgen}(P_H)=\mathsf{OK}(P_G)
      \land P_G\in\mathcal{N}(P_C).
\end{aligned}
\]

The equality comparison covers complete \ccname{Csharpminor} programs, including all
function bodies, global definitions, public symbols, and the program entry
point. The validator therefore cannot accept a candidate merely because an
individual reconstructed function or expression happens to match.

\begin{theorem}[\ccname{Csharpminor}-to-\ccname{Clight} semantic preservation]
Let $P_C$ be a \ccname{Csharpminor} program and $P_H$ an accepted \ccname{Clight}
reconstruction. Then
\[
\begin{aligned}
  &\mathsf{validate}(P_C,P_H)=\mathsf{true}
  \land P_H\Downarrow B_H
  \rightarrow
  \exists B_C.\;
  P_C\Downarrow B_C
  \land B_H\preceq B_C.
\end{aligned}
\]
\end{theorem}

\paragraph{Proof sketch.}
Validator soundness produces a \ccname{Csharpminor} program $P_G$ such that
\ccname{Cshmgen} translates $P_H$ to $P_G$ and
$P_G\in\mathcal{N}(P_C)$. CompCert's
\texttt{Cshmgenproof.transl\_program\_correct} gives a forward simulation from
the semantics of $P_H$ to the semantics of $P_G$.

If $P_G=P_C$, this simulation directly establishes the result. Otherwise,
$P_G$ is one of the permitted normalized programs. Each cleanup and
restoration transformation has a proved reverse simulation relating its
output to its input. We compose the simulations required by the applicable
normalization case to obtain a simulation from $P_G$ to $P_C$. Finally,
composition with the \ccname{Cshmgen} simulation yields a simulation from the
accepted \ccname{Clight} reconstruction $P_H$ to the supplied \ccname{Csharpminor} program
$P_C$. CompCert's generic behavior-preservation theorem then gives the stated
behavior-improvement result.

\begin{corollary}[Preservation of defined \ccname{Clight} behaviors]
\[
\begin{aligned}
  &\mathsf{validate}(P_C,P_H)=\mathsf{true}
  \land P_H\Downarrow B
  \land \mathsf{Safe}(B)
  \rightarrow P_C\Downarrow B.
\end{aligned}
\]
\end{corollary}

The batched discarded-result cleanup has its own semantic-correctness proof,
avoiding repeated whole-program cleanup passes. Consequently, neither Ascent
nor the OCaml candidate generator belongs to the trusted base: an incorrect
type, layout, or declaration inference can make validation fail, but every
candidate accepted by the validator satisfies the theorem above.

\subsection{\ccname{Clight} to \ccname{Csyntax}/Readable C}

The final reconstruction recovers a \ccname{Csyntax} program that can be printed as
recompilable C. It reconstructs source-level expression grouping, loop
initializers, short-circuit expressions, comma expressions, and structured
conditionals. The supplied input $P_L$ is the second \ccname{Clight} form in CompCert's
pipeline: \ccname{SimplExpr} lowers C expressions to \ccname{Clight}, after which
\ccname{SimplLocals} converts eligible local variables to temporaries.

The generator is untrusted. The validator runs these two verified passes on
the candidate $P_C$ and compares the complete generated \ccname{Clight} program with
$P_L$. Exact equality is too restrictive because the forward passes can
introduce administrative \texttt{skip} prefixes and temporary result copies.
Let $S$ remove only the proved skip and guarded-assignment patterns, and let
$R$ fold only proved \texttt{call}/\texttt{builtin} result-copy patterns into
direct result assignments. The validator uses the following bounded family:
\[
\begin{aligned}
\mathcal N=\{\,
  &I,\;S,\;R,\;S\circ R,\;R\circ S,\;S\circ R\circ S,\\
  &R\circ S\circ R\circ S,\;S\circ R\circ S\circ R\circ S
\,\}.
\end{aligned}
\]
The repeated forms are needed because one cleanup can expose an administrative
pattern recognized by the other. The search remains finite, and every
application of $S$ or $R$ has its own proved \ccname{Clight} simulation.

Writing $\mathcal V_C(P_L,P_C)$ for the validator, soundness establishes
\[
\begin{aligned}
  &\mathcal V_C(P_L,P_C)=\mathsf{true}
  \Longrightarrow
  \exists P_{L1},P_G,\;N\in\mathcal N.\\
  &\qquad \mathsf{SimplExpr}(P_C)=\mathsf{OK}(P_{L1})
  \land \mathsf{SimplLocals}(P_{L1})=\mathsf{OK}(P_G)
  \land N(P_G)=P_L.
\end{aligned}
\]
The equality $N(P_G)=P_L$ compares complete \ccname{Clight} programs, including
function bodies, global definitions, public symbols, the main identifier, and
composite-type definitions.

\begin{theorem}[\ccname{Clight}-to-\ccname{Csyntax} semantic preservation]
Let $P_L$ be a \ccname{Clight} program and $P_C$ an accepted \ccname{Csyntax} reconstruction.
Then
\[
\begin{aligned}
  &\mathcal V_C(P_L,P_C)=\mathsf{true}
  \land P_C\Downarrow B_C
  \Longrightarrow
  \exists B_L.\;
  P_L\Downarrow B_L
  \land B_C\preceq B_L.
\end{aligned}
\]
\end{theorem}

\paragraph{Proof sketch.}
Validator soundness produces $P_{L1}$, $P_G$, and $N\in\mathcal N$.
CompCert's correctness theorem for \ccname{SimplExpr} gives a simulation from
$P_C$ to $P_{L1}$, and the theorem for \ccname{SimplLocals} gives a simulation
from $P_{L1}$ to $P_G$. If $N=I$, checked equality identifies $P_G$ with
$P_L$. Otherwise, we compose the simulation from $P_G$ with the cleanup
simulations in $N$. The checked equality $N(P_G)=P_L$ then identifies the
result with the supplied \ccname{Clight} program. Composition therefore gives a
simulation from the accepted \ccname{Csyntax} program $P_C$ to $P_L$. CompCert's
generic behavior theorem produces a \ccname{Clight} behavior $B_L$ satisfying
$B_C\preceq B_L$.

\begin{corollary}[Preservation of defined C behaviors]
\[
\begin{aligned}
  &\mathcal V_C(P_L,P_C)=\mathsf{true}
  \land P_C\Downarrow B
  \land \mathsf{Safe}(B)
  \Longrightarrow P_L\Downarrow B.
\end{aligned}
\]
\end{corollary}
$\mathsf{Safe}(B)$ excludes undefined behavior of the reconstructed C
program. Therefore $B\preceq B_L$ implies $B_L=B$, and the supplied \ccname{Clight}
program has the same behavior $B$.

This theorem closes the source boundary when the preceding
\ccname{Csharpminor}-to-\ccname{Clight} validator supplies the same $P_L$. The complete pipeline
also has a verified replay path for candidates rejected by this local
comparison: it runs \ccname{SimplExpr} and \ccname{SimplLocals}, applies the
same finite cleanup family, and validates the resulting \ccname{Clight} program directly
against the reconstructed \ccname{Csharpminor} input. The subsequent pretty-printer
serializes the accepted \ccname{Csyntax} program and is outside the semantic
transformation.

Composing these pass-local results with the complementary middle-level simulations yields the whole-decompiler theorem stated in Section~\ref{sec:formalization}.

\relax

\subsection{Discharging the C-Safety Assumption}

For a closed checksum recovered without manual cleanup, Rocq checks an
untrusted finite execution graph containing 97 states and 130 edges. The check
covers progress, memory operations, and every C expression-evaluation order,
proving that all executions terminate with an empty trace and result 222. The
composed theorem transfers this result to the supplied \ccname{Asm}. This example shows
that the final C-safety premise can be proved, but does not claim safety for all
reconstructed applications. The artifact records the full search protocols,
certificate, assumption audit, and machine-readable results.

\subsection{Accepted-Chain Signature Fidelity}
\label{sec:accepted-signature-fidelity}

Validator acceptance establishes consistency with the inferred typed
\ccname{Asm}; it does not establish recovery of CompCert's held-out interface.
We therefore compare the complete accepted assignment with the held-out
assignment after validation. On the frozen 15-context zlib campaign, 104/152
(68.4\%) internal-function signatures and 373/449 (83.1\%) call-site signatures
match exactly; only 3/15 contexts have a fully exact signature set.

We also audit every cumulative-portfolio acceptance for which both the archived
receipt and held-out assignment remain available: 1,198/2,606 contexts. In this
receipt-backed subset, 6,809/13,373 (50.9\%) internal-function signatures and
44,074/64,802 (68.0\%) call-site signatures match exactly, and 112/1,198
(9.3\%) contexts are exact throughout. The other 1,408 acceptances lack
comparable archived receipts; we neither classify them nor extrapolate from
the non-random subset.

\subsection{Uniform Cost Ablation}
\label{sec:uniform-cost-ablation}

We replay libuv with one producer, candidate bound, and timeout. Assembly-only
inference accepts 26/37 contexts after 1,878 candidates in 673.468 seconds
(median 7.193, p95 71.327). Adding checked platform, Ghidra, and IDA proposals
accepts 31/37 after 1,365 candidates in 590.192 seconds (median 5.218, p95
90.966). The accepted units contain 258/706 versus 430/706 functions; these are
functions contained in accepted translation units, not independently certified
functions.

\subsection{Signature-Free Execution Audit}
\label{sec:signature-free-execution-audit}

On 30 fixed executable fixtures, inferred interfaces yield 27/30 accepted
chains and 23/30 native passes. Exact interfaces yield 30/30 and 29/30. Of the
four accepted inferred outputs not run natively, three printed C files are not
reparsable by CompCert and one reaches a documented Clang adapter limitation;
the latter also accounts for the exact-interface native failure.

CompCert's interpreter reports 2 expected terminations, 15 internal
stuck/undefined-behavior outcomes, 5 unsupported external calls, and 1
inline-assembly stop on the 23 inferred outputs that pass natively. Exact
interfaces give the same aggregate split, implicating reconstructed body and
layout safety rather than signature inference alone. In a separate current
zlib replay, inference accepts 10/15 translation units. Replacing those ten
objects while retaining five originals passes 3/3 workloads; this hybrid is
not a full signature-free application reconstruction.

\subsection{Native Clang Profile}
\label{sec:native-clang-profile}

The CompCert-compatible Clang profile for recovered C contains four flags:
\begin{compactitem}
\item \texttt{-ffreestanding}
\item \texttt{-ffp-contract=off}
\item \texttt{-fwrapv}
\item \texttt{-fno-strict-aliasing}
\end{compactitem}
The first permits the recovered entry prototype, the second prevents
contraction absent from the CompCert run, and the last two align signed-overflow
and aliasing assumptions with CompCert C.
Ordinary original/native caller code uses \texttt{-ffp-contract=off}; the
receipts preserve the complete command lines and any fixture-specific adapter.

\end{document}